\documentclass{aa}  

\usepackage{amsmath, amsfonts, amssymb}
\usepackage{graphicx,color}
\usepackage[dvipsnames]{xcolor}
\usepackage{txfonts}
\usepackage{bbding}
\usepackage{pifont}
\usepackage{hyperref}
\usepackage{xspace}
\usepackage{subfig}
\usepackage{caption}
\usepackage{rotating}

\newcommand{\hi}{\textsc{H$\,$i}\xspace}
\newcommand{\numunit}[2]{\mbox{\ensuremath{#1\,#2}\xspace}}

\newcommand{\kms}{\ensuremath{\text{km}\,\text{s}^{-1}}\xspace}
\newcommand{\msun}{\ensuremath{\text{M}_\odot}\xspace}
\newcommand{\mk}{MeerKAT\xspace}

\newcommand{\sofia}{\texttt{SoFiA}\xspace}

\newcommand{\hwone}{The Swarm\xspace}
\newcommand{\amod}{\ensuremath{A_{mod}}\xspace}

\graphicspath{{images/}}

\begin{document}

   \title{An \hi\ story of galaxies in Abell 2626 and beyond}

   
   \author{T. Deb,\inst{1,2}\thanks{contact: tirna1106@gmail.com}
          M.A.W. Verheijen,\inst{1} 
          J.M. van der Hulst\inst{1}
          }
\titlerunning{\hi\ story of galaxies in Abell 2626 and beyond}
   \authorrunning{Deb et al.}
   
  \institute{Kapteyn Astronomical Institute, University of Groningen, Landleven 12, 9747 AV Groningen, The Netherlands
  \and
  Department of Physics and Astronomy, University of the Western Cape, Robert Sobukwe Road, Bellville 7535, South Africa}

   \date{Received Sep 7, 2022 / Accepted May 22, 2023}

 
  \abstract
  {To study the effects of environment on galaxies, we use \hi observations of galaxies in and around the cluster Abell 2626 (A2626). The cluster can effectively be divided into three different environments: the cluster itself, a group environment in the periphery of the cluster (we call it the Swarm), and substructure in the cluster itself. We use these to study the dependence of the galaxy properties on environment. }
   {We have explored the relation between \hi deficiency, \hi morphology, and star formation deficiency for the galaxies in and around the A2626 galaxy cluster to investigate the environmental effects on these properties. } 
   {To quantify the asymmetries of the outer \hi disc of a galaxy, we used 1) three visual classes based on the outermost reliable \hi contour (settled, disturbed, and unsettled \hi discs), 2) the offset between the \hi centre and the optical centre of a galaxy, and 3) the modified asymmetry parameter \amod as defined previously. }
   {The \hi deficiency of a galaxy is strongly correlated with the projected distance from the centre of A2626. Furthermore, substructure galaxies tend to be more asymmetric than the isolated galaxies in A2626, probably because tidal interactions are more efficient within the substructures than outside the substructures. Moreover, asymmetric, offset, and smaller \hi discs are not necessarily the result of the cluster environment because they are also observed in substructures in A2626 and in the Swarm. This signifies that a pre-processing of the \hi discs of galaxies in groups or substructures plays an important role, together with the processing in the cluster environment. Finally, the star formation rates for the galaxies in all three environments are slightly lower than the typical star formation rate for normal galaxies, as manifested by their offset from the star formation main sequence. This implies effective gas removal mechanisms in all three environments.}
  {} 
  


   \maketitle
%

 \section{Introduction}
   \label{sec:intro}

The well-known morphology-density relation \citep{Dressler1980} provides direct evidence of the notion that the cosmic environment of galaxies influences their constitutional properties and star formation activity both during their formation (nature) and in the subsequent evolution and interaction with their surroundings (nurture). In the hierarchical structure formation scenario \citep{Toomre1972}, galaxy clusters grow by accreting galaxies, galaxy groups, and other clusters along the filaments of the cosmic web. Diverse astrophysical processes impact the morphologies, gas content, and star formation properties of the galaxies when they move into the dense cluster environment. Within $R_{200}$ ($\approx 0.75 R_{vir}$) of a cluster, gravitational perturbations such as tidal galaxy-galaxy interactions and mergers (e.g. \citealt{Spitzer1951,Tinsley1979,Merritt1983,Springel2000}), tidal galaxy-cluster interactions \citep{Byrd1990,Valluri1993}, and galaxy-galaxy interactions (sometimes called galaxy harassment; e.g. \citealt{Moore1996, Jaffe2016}), affect both the stellar and gaseous components of the galaxies. Hydrodynamical processes such as starvation (e.g. \citealt{Larson1980, Balogh2000}), thermal evaporation \citep{Cowie1977}, ram-pressure stripping (e.g. \citealt{Gunn1972}), and viscous stripping \citep{Nelson1982} only affect the gas content of galaxies. Moreover, these processes are sometimes effective in transporting cold gas to the centrs of galaxies and may trigger Active Galactic Nucleus (AGN) activity \citep{Baldry2004, Balogh2009, Poggianti2017}.

While neutral atomic hydrogen or \hi gas in galaxies provides the raw fuel for star formation, the morphologies and kinematics of the extended, collisional, and fragile \hi gas discs serve as sensitive diagnostic tracers of the environmentally driven processes (e.g. \citealt{Davies1973, Haynes1984, Williams1987, Abramson2011, Serra2013, Jaffe2015}). A good illustration of \hi in galaxies in a cluster environment was provided by the Very Large Array (VLA) Imaging survey of Virgo galaxies in Atomic gas (VIVA; \citealt{chung2009}): Galaxies near the core of the Virgo cluster show small or disturbed \hi discs with a low column density, while in many cases, the \hi gas is displaced from the stellar body, trailing the galaxy along its infall trajectory, and providing evidence of ram-pressure stripping. Occasionally, star formation occurs \textit{\textup{in situ}} in these gas tails, ionising the gas and giving rise to the jellyfish phenomenon (e.g. \citealt{Poggianti2017, ramatsoku2019, ramatsoku2020, Deb2020}). In two other nearby clusters, Coma (e.g. \citealt{Molnar2021}) and Fornax \citep{loni2021}, at least half of the \hi detections show a disturbed \hi morphology, including several cases of asymmetries, tails, offsets between \hi and optical centres, and truncated \hi discs. While most of the \hi-selected Coma galaxies have enhanced star formation rates and are also \hi deficient compared to field galaxies of the same stellar mass \citep{Molnar2021}, Fornax galaxies are \hi deficient and have a low star formation rate \citep{loni2021}. This means that the cluster environment affects both the \hi gas content and the star formation rate in galaxies. The extent of the disturbance of the \hi morphologies and star formation depends on the properties of the cluster environment and the galaxies. However, \hi deficient and quenched galaxies are also observed in less dense environments, including cluster outskirts, galaxy groups, and the filaments of the cosmic web, both in observations and simulations (e.g. \citealt{Solanes2001, Tonnesen2007, Yoon2017}). This means that the galaxies may already be pre-processed, that is, they have lost their \hi gas and their star formation activity was halted prior to their entering the dense cluster environment. It is important to understand the nature and efficiency of pre-processing that occurs at larger distances from the cluster centres, beyond $R_{200}$, where galaxies can be relatively isolated or reside in groups with miscellaneous dynamic histories and galaxy populations (e.g. \citealt{Laigle2018, Kraljic2018, Alpaslan2016, Sarron2018, Kleiner2017, Vulcani2019}). Wide-area volume-limited \hi imaging surveys of galaxy clusters that probe the galaxies beyond the virial radius of the cluster are therefore indispensable for investigating different environmental mechanisms in different clusters for a better understanding of the impact of the different environments on the galaxies.

\begin{figure}[ht!]
 {
    \includegraphics[width=0.5\textwidth]{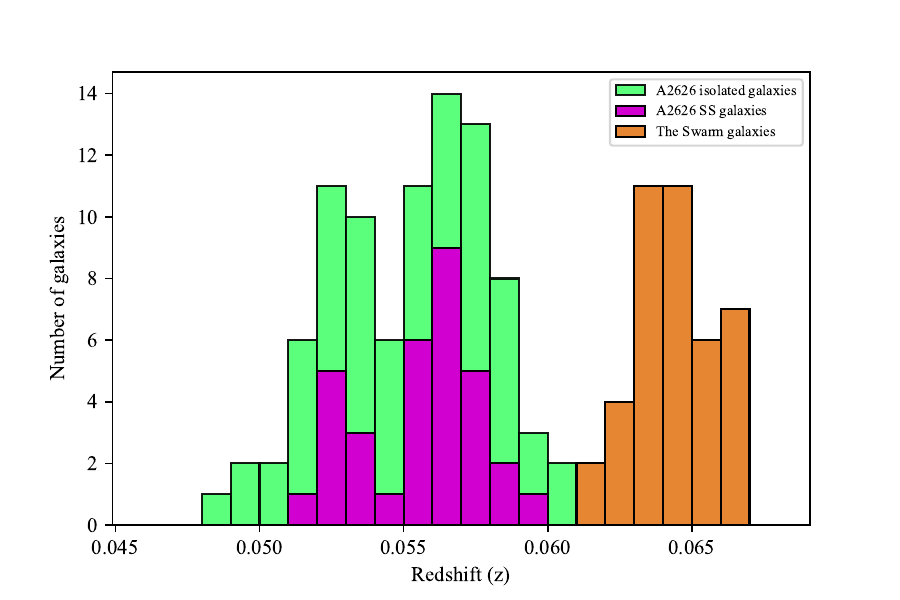}
  } 
  \caption{Redshift distribution of three different environments in and around A2626. The green, magenta, and orange histograms present non-substructure and substructure galaxies in A2626 and galaxies in the Swarm, respectively. } 
  \label{fig:A2626_TS_histogram} 
 \end{figure}

However, it is difficult to identify the ongoing astrophysical mechanisms for individual galaxies within an observed volume around a cluster from their \hi morphologies alone. Moreover, if the galaxies are not well resolved, it is even difficult to unambiguously classify their \hi morphologies. Although it is not possible to confirm the corresponding environmental mechanisms in detail by only investigating the \hi morphologies of the galaxies, it is still possible to broadly identify the plausible environmental mechanisms acting on galaxies based on their \hi morphologies. Moreover, when we also consider \hi deficiencies, star formation rates, optical colours and morphologies, and the locations of the galaxies with respect to the cluster core together with the \hi morphologies, we would gain further insight into the corresponding astrophysical interactions with the environment. 

In this work, we explore the \hi deficiencies and morphologies of galaxies and relate them to the corresponding environmental mechanisms for the \hi detected galaxies in and around the A2626 cluster, which was recently observed with \mk \citep{HD2021}. For this purpose, we employ different methods for classifying or quantifying \hi morphologies  to understand which method is most suitable for identifying a particular environmental process. 

\section{The Sample}
\subsection{A2626 and its neighbourhood}

\label{subsec:diff_envs}

 A2626 is a fairly massive  ($\sim$ 5 $\times$ 10$^{14}$ \msun) galaxy cluster located at a  redshift of z=0.055 \citep{HealySS2021}.  With \mk, we surveyed a large cosmic volume that contains three distinct over-densities that represent three different global environments. The galaxies within the A2626 cluster are in the redshift range $0.0475 < z < 0.0615$. Behind A2626, at the redshift range $0.0615 < z < 0.0675$, we find a collection of galaxy groups that we refer to as the Swarm. Behind the Swarm, we find another over-density in the redshift range $0.0675 < z < 0.0745$ that is associated with the cluster A2637 (see Fig. 6 and 11 in \citealt{HD2021} and Fig. 5(a,b,c) in \citealt{HealySS2021} ). This cluster is located north-east of A2626, near the full-width-half-maxima (FWHM) of the primary beam of \mk. Its galaxies are therefore less well visible because there is significant primary-beam attenuation. Therefore, we only considered galaxies at the redshifts in the over-densities of A2626 and \hwone (see Fig. \ref{fig:A2626_TS_histogram}). Table \ref{tab:cluster_properties} presents the different properties of A2626 and the Swarm. \citet{HealySS2021} have identified six different substructures in the A2626 cluster based on the  Dressler-Shectman (DS) test (see Fig. 13 in \citealt{HealySS2021}). We note that in one of the substructures, close to the cluster centre,  none of the galaxies are detected in \hi. Because only a relatively low number of \hi detections are made in each substructure, we combined the \hi detections of all the substructures together in order to analyse the \hi properties of the substructure galaxies. Within these two over-densities (A2626 and the Swarm), we identify three classes of galaxies, depending on their environment.
  
  \renewcommand{\arraystretch}{1}
\begin{table}
    \centering
    \caption{Properties of A2626 and the Swarm.}
    \begin{tabular}{lclc}\hline 
   Environment & cz ($km/s$) & $\sigma$ ($km/s)$ & R$_{200}$ (Mpc)                 \\ \hline 
 A2626  & 16576 &660 $\pm$ 26  & 1.59\\
 The Swarm   & 19247 & 397 $\pm$ 22 & 0.95\\
     \hline
    \end{tabular}
    \label{tab:cluster_properties}
\end{table}

\begin{enumerate}
    \item  The first class consists of galaxies in A2626 that lack substructure or are isolated. Fifty-seven galaxies lie in the cluster environment, but are not a member of a substructure. However, we note that these non-substructure galaxies still lie in a high-density environment.
    \item  The second class consists of substructure galaxies in A2626: Thirty-four galaxies that are within the substructures in A2626. They are in groups within the cluster environment. 
    \item  The final class consists of the Swarm galaxies. Thirty-two galaxies that are within the group environment associated with the Swarm, which is a structure behind the cluster.
\end{enumerate}

 \noindent We explore the environmental impact on the \hi deficiencies and morphologies of these 122 galaxies, that might be processed due to the cluster environment, or pre-processed before they fall into the cluster.
 
 \subsection{Description of available data products}
 
 A detailed description of the \hi observations, data reduction, and \hi data products is provided in \cite{HD2021}. We have \hi maps, \hi signal-to-noise ratio maps, velocity fields, position velocity diagrams, and \hi global profiles for all 219 \hi detections in the entire surveyed volume (see \citealt{HD2021} for details). The sensitivity of MeerKAT enables the detection of \hi emission well beyond the FWHM of the primary beam, and we therefore imaged an area of 2 $\times$ 2 deg$^{2}$ centred on A2626. For this work, we analysed the \hi data at an angular  resolution of 15" , which corresponds to a spatial resolution of $\sim$15 kpc at the distance of A2626. At the centre of the field of view and at the redshift of A2626, we reach a $3\sigma$ column density sensitivity of \numunit{\mathrm{N}_\hi = 4.6 \times 10^{19}}{\text{cm}^{-2}} per channel (207 kHz wide, corresponding to 45 km s$^{-1}$) at a resolution of $15''$  (see Figure 4 in \citealt{HD2021}).
 
Photometric imaging data were taken from the DECam Legacy Survey \citep{dey2019} and from the Sloan Digital Sky Survey (SDSS; \citealt{York2000, Aguado2018}). Optical spectroscopic data were mainly taken from the SDSS and a new dedicated MMT/ Hectospec survey \citep{HealySS2021}. Stellar masses and star formation rates were calculated from WISE photometry that was kindly provided by T. Jarrett (private communication). Stellar masses were calculated from W1$(3.4 \mu$m) and W2 (4.6$\mu$m) fluxes, and the SFRs are based mainly on the 12 $\mu$m emission.

For our analysis, we adopted optical centres from the SDSS, and we measured the \hi centres by fitting 2D Gaussians to the \hi maps of the galaxies. To quantify how well an \hi map is resolved, we divided the number of pixels above a certain column density level in the \hi map (e.g. the column density corresponding to three times the signal-to-noise ratio) by the number of pixels in the \mk synthesised beam. 

\bigskip The remainder of the paper is structured as follows: Sec. \ref{sec:hi_morphologies} presents different ways of classifying \hi morphologies. In Sec. \ref{sec:HI_def} we explore \hi deficiencies of galaxies in A2626 and the Swarm. In Sec. \ref{sec:sfr_dep_time} we investigate the star formation rates and depletion times in the A2626 and the Swarm galaxies. In Sec. \ref{sec:interesting_gals} we show some interesting \hi morphologies of selected galaxies. In Sec. \ref{sec:discusiions} we discuss our observational findings and address the questions raised by our results. Finally, in Sec. \ref{sec:summary}, we summarise our work. We have used \numunit{{H}_0 = 70}{\kms}, $\Omega_{m} = 0.3$, and $\Omega_{\Lambda} = 0.7$.
  
\section{Classifying \hi morphologies}
 \label{sec:hi_morphologies}

\hi asymmetries are thought to be indicative of environmental effects. We aim to study them in relation with the local environment and then explore which method(s) is robust enough to assess \hi morphologies in the A2626 galaxies given the limitations of the available \mk data (good sensitivity that variable across the field of view, however, and moderate linear and spectral resolution). To be able to do this, we first need to examine the detected sources carefully and prune the sample so that only the objects with sufficient sensitivity and resolution are considered for further analysis. 

\begin{figure}[t!]
\begin{center}
{\includegraphics[width=0.45\textwidth]{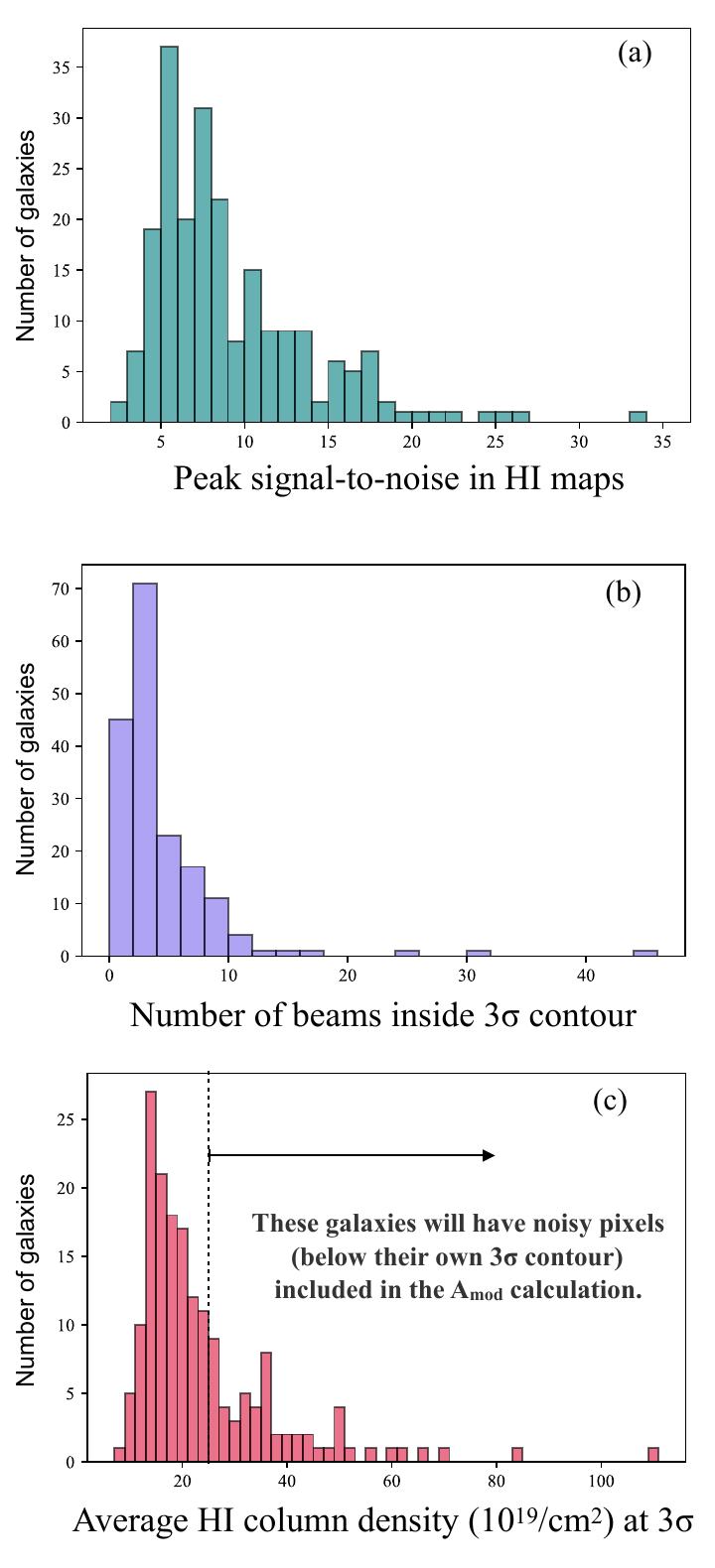}} 
  \caption{Histograms of three observational properties of galaxies in and around A2626. (a) Distribution of the peak signal-to-noise ratio in the \hi maps. (b) Number of beams enclosed by the 3$\sigma$ contours of the \hi maps. (c) Distribution of 3$\sigma$ column density levels in the \hi maps. The 3$\sigma$ \hi column density used here is the average column density from the signal-to-noise ratio map where the pixels with 2.5< (S/N) < 3.5 were selected following \citet{HD2021}.}
  \label{fig:S2N_histogram} 
  \end{center}
 \end{figure}

In the observed volume, 219 \hi sources identified by \sofia \citep{Serra2015} have optical counterparts (see \citealt{HD2021}). Among these 219 sources, we first identified the uncertain \hi detections with a low signal-to-noise ratio that have only a few pixels with a very high signal-to-noise ratio, or sources that are missing one peak of the double-horn profile in the \sofia mask, as revealed in the position-velocity diagram, and excluded them from our analysis. We then were left with 177 reliable \hi sources with a peak signal-to-noise ratio in their \hi map $\geq$ 5, as indicated by the signal-to-noise ratio maps in the atlas pages in \citet{HD2021}. One hundred and eight of these 177 galaxies lie in the A2626 cluster and ine the Swarm. The number of \hi sources in each of these environments is given in Table \ref{tab:hi_detections}. In the top panel of Fig. \ref{fig:S2N_histogram}, we present a histogram with the distribution of the peak signal-to-noise ratio values for these 177 galaxies. Therefore, this histogram gives an impression of the quality of the data. The median peak signal-to-noise ratio of the galaxies in the surveyed volume is 8.6. The middle panel of Fig. \ref{fig:S2N_histogram} shows the histogram of the number of beams within the 3$\sigma$ contour of the \hi map (see \citealt{HD2021} for details). The median number of beams for these 177 galaxies is 3.1.

\renewcommand{\arraystretch}{1.1}
\begin{table}
    \centering
    \caption{\hi sources in different environments.}
    \begin{tabular}{lc}\hline
   Environment & No. of \hi sources                   \\ \hline
  Non-substructure in A2626  & 46\\
Substructures in A2626  & 30\\
 The Swarm  & 32\\
     \hline
    \end{tabular}
    \label{tab:hi_detections}
\end{table}

\subsection{Visual classifications}

\label{subsec:visual_classes}

We first visually classified galaxies in our entire sample based on their \hi morphologies. For the visual classification, we assessed the \hi morphologies based on the 3$\sigma$ column density contours and considered three different classes: settled, one-sided or disturbed, and unsettled \hi discs, similar to \cite{Molnar2021}, who classified \hi morphologies of the galaxies in the Coma cluster.  

\begin{enumerate}
    \item \textbf{Settled sources} (Vclass 1) are sources in which the \hi distribution is already settled, that is, their symmetric \hi morphologies are centred on a stellar disc and their  velocity gradient is consistent with rotation.  Spatially unresolved \hi sources were also included in this category.

\item \textbf{Disturbed sources} (Vclass 2) are \hi sources with either a regular disc component with an additional one-sided asymmetry in their \hi morphology, or whose \hi distribution is regular with a significant excess of \hi flux on one side of the stellar disc.

\item \textbf{Unsettled sources} (Vclass 3) are \hi sources with an irregular \hi morphology and/or kinematics or with \hi flux extensions in multiple directions from the stellar disc. They can also have an extreme 3D asymmetry or displacement relative to the
optical light with an unclear \hi disc component.

\end{enumerate}

Table \ref{tab:hi_visual_classes} shows the number of galaxies with different visual classifications in different environments. Thus, substructure galaxies are more disturbed or unsettled than non-substructure galaxies in A2626 or the Swarm. Although the statistics is limited for the galaxies in these three environments, there are significantly more ($\sim$ 84\%) disturbed or unsettled galaxies in substructures than among the non-substructure galaxies (65\%) in A2626 or in the Swarm  galaxies (60\%). In substructures, only 17\% $\pm$ 8\% of the galaxies have settled \hi discs, while in the other two environments, 37\% $\pm$ 8\% of the galaxies have settled \hi discs. This difference is at the 2$\sigma$ level. 

\renewcommand{\arraystretch}{1.4}
\begin{table*}
    \centering
    \caption{Number of galaxies with different visual classifications in different environments.}
    \begin{tabular}{lccc}\hline 
   Environment & No. of & No. of  & No. of   \\ 
   & VClass 1 sources & VClass 2 sources & VClass 3 sources\\     \hline \hline
  non-substructure in A2626  & 16 (35\%) & 25 (54\%) & 5 (11\%)\\
Substructures in A2626   & 5 (16.7\%) & 23 (76.7\%)& 2 (6.6\%) \\
 The Swarm   & 13 (40\%) & 12 (38\%) & 7 (22\%) \\
     \hline
    \end{tabular}
    \label{tab:hi_visual_classes}
\end{table*}

\subsection{\hi offsets}

\begin{figure}[t]
\centering
 {\includegraphics[clip, trim=0cm 0cm 0cm 0cm, width=0.5\textwidth]{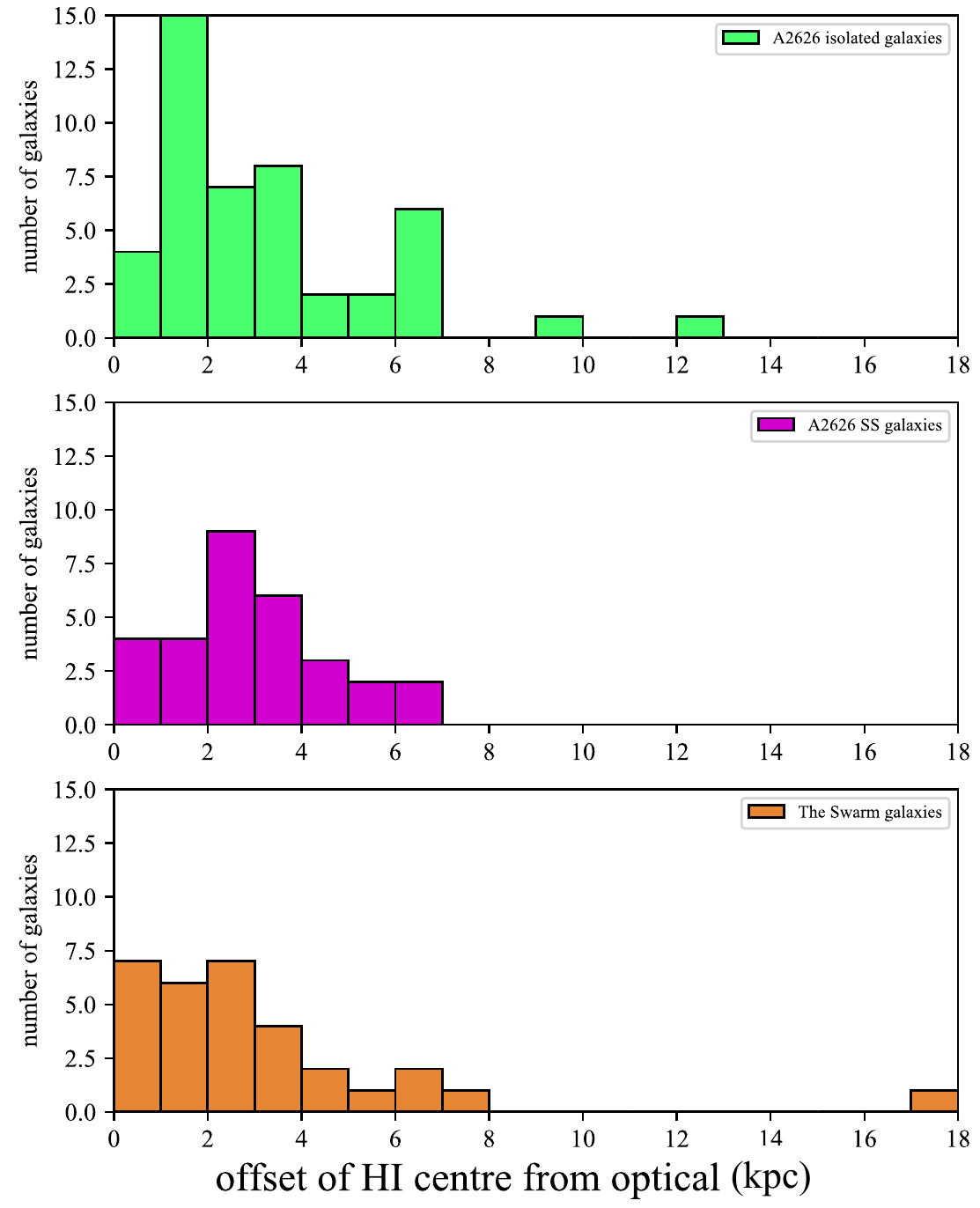}} 
  \caption{Distribution of \hi offset in kiloparsec presented as histograms for the galaxies in and around A2626. The top, middle, and bottom panels show \hi offsets for the non-substructure (green) and substructure (magenta) galaxies in A2626, and for the galaxies in the Swarm (orange), respectively.} 
  \label{fig:offset_histogram} 
 \end{figure}

Another effective way to quantify the asymmetry in the \hi distribution is to measure the offset of the \hi centre from the optical centre. \hi centres are calculated by fitting 2D Gaussians to the \hi maps while optical centres are taken from the SDSS.
A significant offset of the \hi centre from the optical centre may signify an external environmentally induced disturbance in the \hi morphology. We note that even with a small \hi offset, a galaxy may still experience subtle environmental processes such as thermal evaporation or starvation, or it can be at a late stage of ram-pressure stripping. Fig. \ref{fig:offset_histogram} presents the distributions of \hi offset for non-substructure, substructure, and the Swarm galaxies. All the galaxies in these three environments display a range of \hi offsets, signifying that there is no obvious correlation of \hi offset with the different environments in and around A2626.

\subsection{Quantifying asymmetries}

\textbf{Defining the \hi asymmetry}
\newline \noindent
To investigate the environmental impact on the \hi discs of galaxies, it is important to quantify the severity of the asymmetry/ disturbance in the outer \hi distribution of individual galaxies. The optical/ infrared morphologies of galaxies are generally quantified with the parameters concentration, asymmetry, and smoothness (CAS) \citep{conselice2003} and the Gini and M20 parameters \citep{lotz2004}. \cite{holwerda2011, holwerda2013} used these parameters to quantify the \hi morphologies of galaxies from multiple \hi surveys (WHISP, LITTLE-THINGS, and VLA-ANGST). In addition, \cite{holwerda2011, holwerda2013} and \cite{giese2016} used the definition of asymmetry from \cite{conselice2003}. \cite{giese2016} found that the asymmetry parameter is much more robust than any of the other parameters for \hi imaging data,
\begin{equation}
    A =  \frac{\sum_{i,j}^{N}|{I(i,j) - I_{180}(i,j)}|}{\sum_{i,j}^{N}|{I(i,j)}|}
,\end{equation}
where \textit{I(i, j)} denotes the value of the pixel at the \textit{i, j} position of the original image of the galaxy, and  $I_{180}(i, j)$ is the value of the pixel in the same position of the image rotated by 180$^{\circ}$ around the centre of the galaxy. This means that asymmetry is measured by summing the pixel-by-pixel difference of the original and rotated image and normalising this by the total intensity in the image. Hence, the asymmetry index can have a value between 0 and 1. Asymmetries in the fainter outer regions of the \hi discs contribute very little to the global asymmetry index compared to the brighter ($\sim$ 2 orders of magnitude or more) inner regions. Environmental processes, however, influence the extended outer parts of the \hi disc in a galaxy much more easily than the inner parts. In order to give proper weight to the asymmetries in the outer \hi disc, \cite{Lelli2014} introduced a modified asymmetry index ($A_{mod}$),

\begin{equation}
    A_{mod} = \frac{1}{N}\sum_{i,j}^{N} \frac{|{I(i,j) - I_{180}(i,j)}|}{|{I(i,j) + I_{180}(i,j)}|}
,\end{equation}
where $N$ is the total number of pixels in the image. Thus, this definition of $A_{mod}$ normalizes the intensity differences at the position $(i, j)$ with the local intensity of the pixels, in contrast to the total intensity of all the pixels in the \hi map. For example, for a highly lopsided galaxy with \hi emission exclusively from one side, $A_{mod}$ will obtain the maximum value of 1. However, we recognise that the value of \amod depends on some selection criteria.

(i) The value of \amod depends on the \hi column density above which it is measured. Depending on the observational setup and on the sensitivity of the telescope, different \hi observations have different column density sensitivities. Moreover, even with the same observational settings, galaxies may have diverse 3$\sigma$ column density thresholds, depending on the local noise in the vicinity of the galaxy and on the location of the galaxy with respect to the pointing centre, which affects the local primary beam attenuation. 

(ii) The choice of the galaxy centre around which the \hi disc is rotated also affects the measurement of \amod. For disturbed and unsettled \hi distributions, the \hi centre often does not coincide with the optical centre. Moreover, the determination of the optical or the \hi centre also depends on the adopted calculation method. 

(iii) The reliability of the \amod measurement also depends on how well resolved the galaxy is, both in terms of the beam size in kiloparsec and in terms of the number of beams across the \hi map. If the galaxy is only marginally resolved and detected at a low signal-to-noise ratio, the asymmetry in the outer \hi disc might be dominated by noise and not be induced by environmental processes.

\cite{bilimogga2022} have investigated the dependence of \amod on the signal-to-noise ratio, the column density threshold, and the angular resolution of the \hi observations. Using mock galaxies from the EAGLE simulations \citep{schaye2015, crain2015}, they suggested an optimal combination of the observational constraints that are required for a robust measurement of the \amod value of the outer \hi disc of a galaxy: a column density threshold of 5 $\times$ 10$^{19}$cm$^{-2}$ or lower at a signal-to-noise ratio of at least 3, and a galaxy resolved with at least 11 beams. These are not hard limits because they depend on what is considered to be acceptable deviations of the measured \amod from the intrinsic \amod of a galaxy. Our observations do reach this column density sensitivity, but most galaxies are not resolved by 11 beams. 

\medskip

\noindent
\textbf{Measuring the asymmetry in galaxies in and around A2626}

\label{subsec:measuring_hi_asym}
\noindent
To compute \amod, we needed to adopt the position of the centre of the galaxy as well as a column density threshold above which we would consider the intensity of the \hi emission. All 219 \hi detected galaxies in our survey have optical counterparts within the footprints of the SDSS (\citealt{York2000, Aguado2018}) and the DECam Legacy Survey (DECaLS, \citealt{dey2019}). For our analysis, we adopted the optical centres from the SDSS for all 219 galaxies. We did not consider the \hi kinematic centres because most of the galaxies in our sample have disturbed/ unsettled \hi discs. Moreover, optical centres are generally better tracers of the dynamical centre of a galaxy than the \hi centres. When the optical centre is used, the \amod value for the galaxies with a strong offset between the \hi distribution and the stellar body is sometimes high, which may indicate a recent interaction or accretion event. 

It is crucial to calculate \amod above a specific column density for all the galaxies in the sample to make fair comparisons, but this specific column density will correspond to different levels of the signal-to-noise ratio for different galaxies, depending on the local noise properties and on the location of the galaxy within the primary beam. The range of 3$\sigma$ column density sensitivities is (9-65) $\times$ 10$^{19}$ cm$^{-2}$ in our \mk observations \citep{HD2021}. To measure \amod reliably, we need images with an adequate signal-to-noise ratio, and we need the lowest possible column density limit to measure the outer parts of galaxies well. The bottom panel of Fig. \ref{fig:S2N_histogram} shows the distribution of the 3$\sigma$ \hi column density levels for the reliable \hi detections with a peak signal-to-noise ratio $\geq$ 5 in the \hi maps (177 of the 219 galaxies). 

The vertical dotted line in the bottom panel of Fig. \ref{fig:S2N_histogram} represents the \hi column density level of 25 $\times$ 10$^{19}$ cm$^{-2}$.  The 3$\sigma$ \hi column densities of the galaxies to the left of this dotted line are lower than 25 $\times$ 10$^{19}$ cm$^{-2}$. Hence, their \amod values will be reliable when pixels above 25 $\times$ 10$^{19}$ cm$^{-2}$ are included in the calculation of \amod. The 3$\sigma$ \hi column density levels of the galaxies to the right of the dotted line
are higher 25 $\times$ 10$^{19}$ cm$^{-2}$, which means that noisy pixels (below their own 3$\sigma$ contour) are included in the \amod calculation when pixels with values as low as 25 $\times$  10$^{19}$ cm$^{-2}$ are included. By adopting a threshold of 25 $\times$ 10$^{19}$, we retained 71\% of the galaxies for which the \amod calculation only includes pixels with a signal-to-noise ratio >3 in the \hi map. Hence, we considered the corresponding \amod values as reliable, provided the galaxy is sufficiently resolved (see below).

In a statistical sense, for a sample of galaxies with random inclinations, 25 $\times$ 10$^{19}$ cm$^{-2}$ motably corresponds to 1 \msun/pc$^{2}$ in a face-on orientation, which is the typical column density at which the diameters of \hi discs are measured. The 1 \msun/pc$^{2}$ surface density corresponds to a \hi column density of 12.5 $\times$ 10$^{19}$ cm$^{-2}$. Because the galaxies in our sample cover the full range of inclinations, we do not always consider the \hi column density perpendicular to the plane of these galaxies by considering
the \textit{\textup{observed}} column density along the line of sight. Because most of the galaxies in our survey are barely resolved, we cannot make meaningful inclination corrections for the \hi column densities of individual galaxies. Deep optical imaging is often lacking, and for these systems, the inclinations are difficult to determine in general. We therefore adopted a statistical approach. The median inclination of a randomly oriented sample of galaxies is i = 60$^{\circ}$. This means that a face-on column density of 12.5 $\times$ 10$^{19}$ cm$^{-2}$ increases to a line-of-sight column density of 12.5/cos(60$^{\circ}$) = 25 $\times$ 10$^{19}$ cm$^{-2}$. Thus, 25 $\times$ 10$^{19}$ cm$^{-2}$ is also the practical level we adopted for measuring \hi diameters.

Another important aspect for a reliable \amod measurement is the angular resolution. For poorly resolved galaxies, \amod is not very meaningful. In addition to the signal-to-noise ratio, we therefore need to restrict the sample to sufficiently resolved galaxies. We considered the following criteria to identify the galaxies for which the \amod value is reliable: galaxies with a 3$\sigma$ column density level of $\leq$ 25 $\times$ 10$^{19}$ cm$^{-2}$, with a peak signal-to-noise ratio of $\geq$5, and galaxies that are resolved by more than 3 beams. Consequently, the calculation of \amod is meaningful for 33 galaxies. 

\begin{figure}[t!]
 {
    \includegraphics[width=0.45\textwidth]{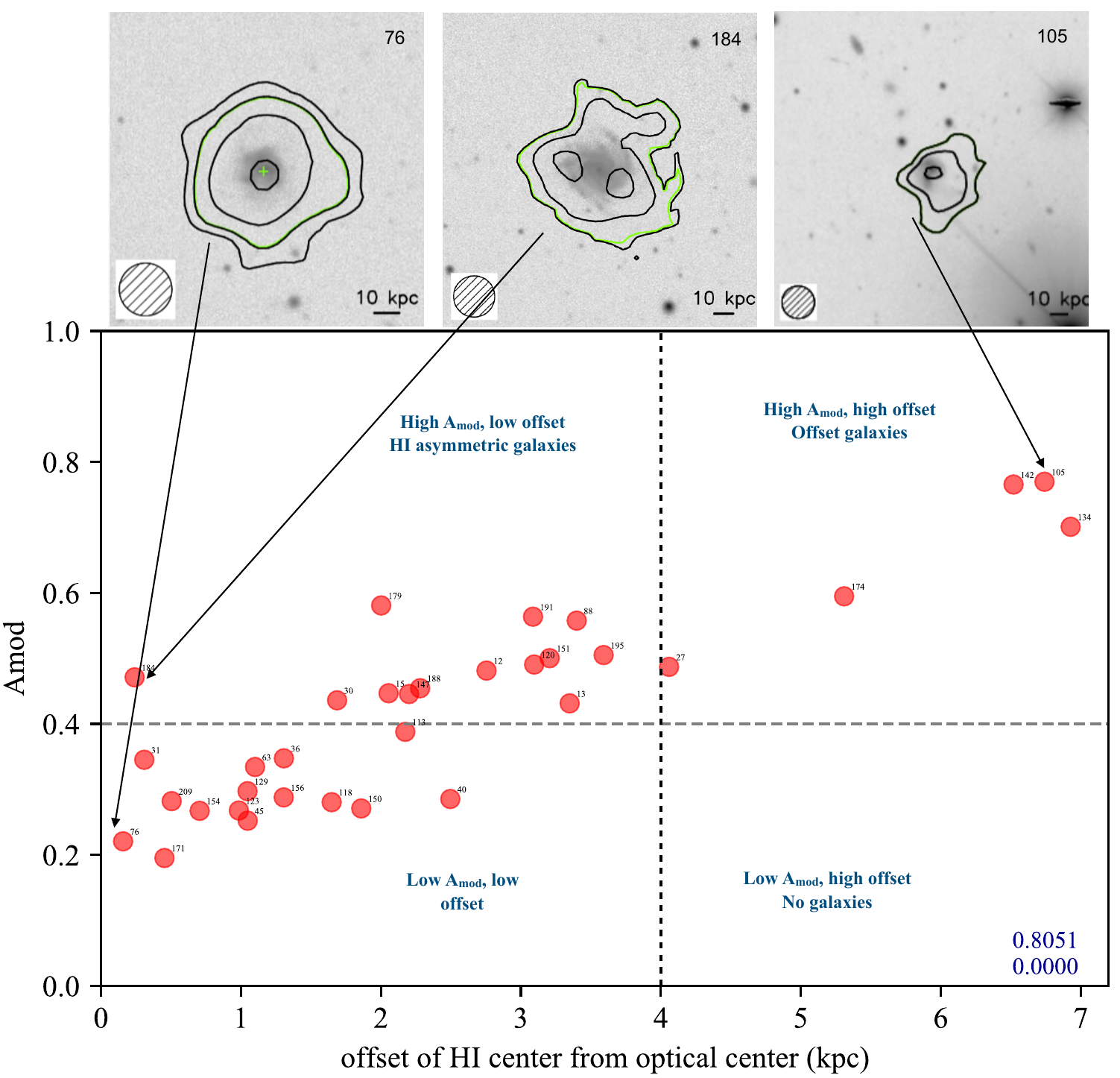} 
  } 
  \caption{Distribution of \amod and \hi offset for the galaxies in and around A2626. The galaxies above the horizontal dashed line (\amod>0.4) are considered as \hi disturbed or unsettled galaxies. Galaxies on the right side of the vertical dashed line are considered as high \hi offset galaxies. \hi maps of a few example galaxies (see Section \ref{subsec:measuring_hi_asym}) are included as insets in the top panel. In the bottom right corner, we reported the Pearson cofficient and p-value for the relation between \amod and the \hi offset.} 
  \label{fig:Amod_offset_class1} 
 \end{figure}

\medskip

\noindent
\textbf{The relation between visual classifications, offsets, and \amod}

\noindent
We explored the relation between our different methods of classifying \hi morphologies: visual classes, \hi offset, and \amod. Fig. \ref{fig:Amod_offset_class1} shows \amod as a function of offset between the \hi and the optical centres. The correlation between \amod and the offset for these 33 galaxies is strong. The Spearman correlation coefficient is 0.81, which is shown in the bottom right corner of Fig. \ref{fig:Amod_offset_class1}. This correlation is expected because a high offset between the \hi and optical centres would result in a high \amod value, but a galaxy with a high \amod value does not necessarily have a strongly offset \hi disc. For practical purposes, we classified a galaxy as \hi asymmetric when its \amod value is $\geq$0.4. Such an \hi asymmetric galaxy will be located above the horizontal dashed line in Fig. \ref{fig:Amod_offset_class1}. Similarly, we defined galaxies to the right of the vertical dashed line in Fig. \ref{fig:Amod_offset_class1} as the high \hi offset galaxies.

These thresholds are indicated by the horizontal and vertical dashed lines. They separate the galaxies into different classes. In particular, we found three different areas in the plot that represent three different types of galaxies. 
\begin{enumerate}
    \item In the upper left quadrant (q1) lie high \amod, low \hi offset galaxies. For these galaxies, a high value of \amod is driven by the asymmetry in the outer \hi disc. These are therefore \hi asymmetric galaxies. For example, in Fig. \ref{fig:Amod_offset_class1}, the asymmetry in the outer \hi contour of galaxy 184 results in a high \amod value.

    \item In the upper right quadrant (q2) lie high \amod, high \hi offset galaxies. For these galaxies, a high \amod is driven by the high \hi offset (sometimes due to the combination of an \hi offset and an outer disc \hi asymmetry). For example, galaxy 105 in  Fig. \ref{fig:Amod_offset_class1} has a strongly offset \hi disc with respect to the optical centre.
     
    \item In the lower left quadrant (q4) lie low \amod, low \hi offset galaxies. The \hi disc of these galaxies is neither asymmetric nor offset. Hence, these are \hi normal galaxies (e.g. galaxy 76 in Fig. \ref{fig:Amod_offset_class1}), although their \hi discs can be small with respect to the stellar disc.

\end{enumerate}

\noindent We note that no galaxies lie in the bottom right quadrant (q3), that is, galaxies with low \amod and high \hi offset. This is expected becaue a high \hi offset would automatically cause a high \amod value. 

\begin{figure*}[t!]
 {
    \includegraphics[width=1\textwidth]{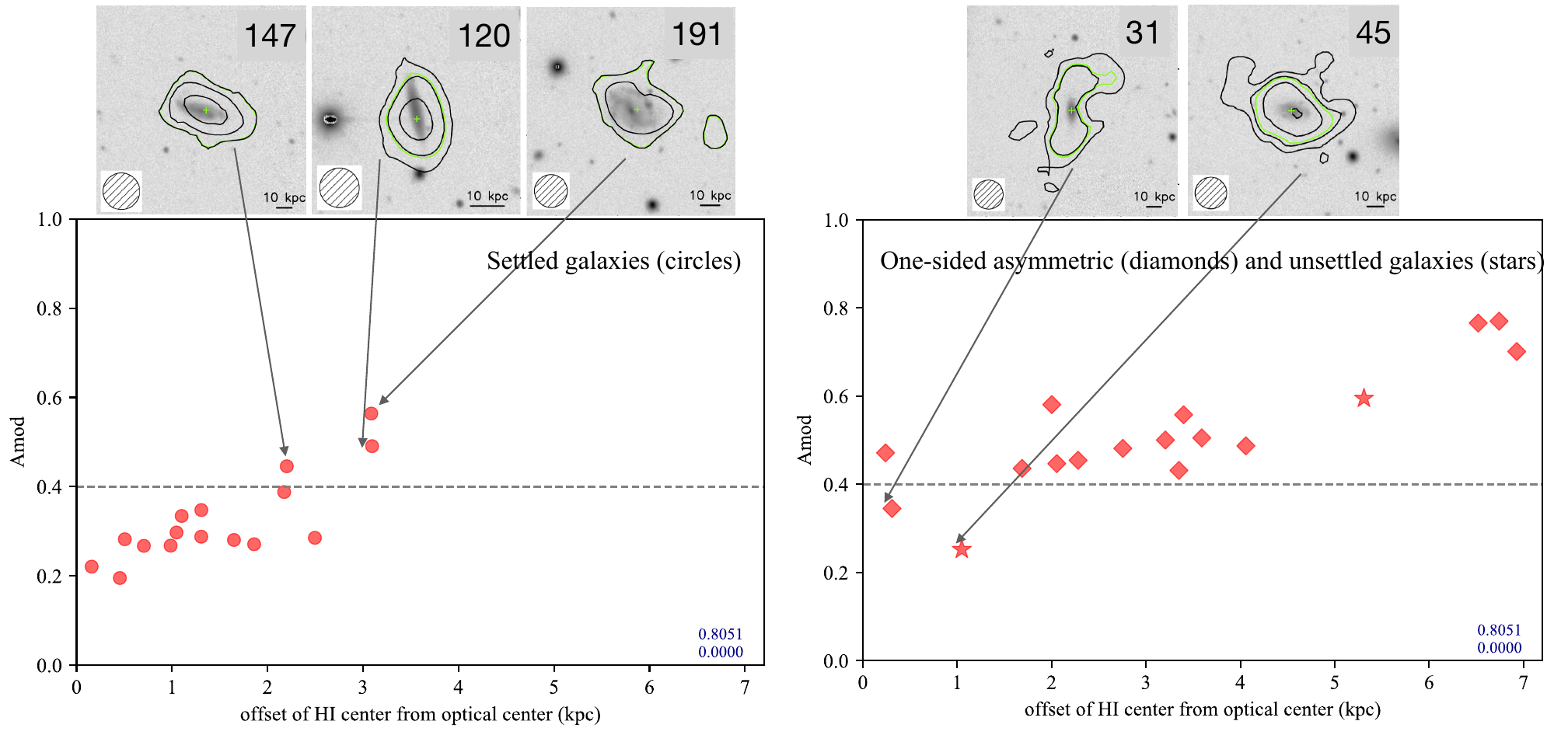} 
  } 
  \caption{Distribution of \amod and \hi offset as a function of visual classes for the galaxies in and around A2626. Left panel: Settled galaxies (circles). The top panels show \hi maps of settled galaxies with high \amod. Right panel: Disturbed (diamonds) and unsettled (stars) galaxies. The top panels show \hi maps of disturbed and unsettled galaxies with low \amod values.   } 
  \label{fig:Amod_offset_visual_class1} 
 \end{figure*}
Next, we explored the same offset-\amod relation with our visual classifications of \hi asymmetries as an additional parameter. The left and right panel of Fig. \ref{fig:Amod_offset_visual_class1} show the galaxies with settled \hi discs as circles and the galaxies with disturbed / unsettled \hi discs as diamonds and stars, respectively, based on our visual classifications as described in Sec. \ref{subsec:visual_classes}. When our visual classification completely overlaps the \hi asymmetries based on the \amod values, all the diamonds and stars should therefore be above the \amod = 0.4 horizontal dashed line and all the circles should be below the \amod = 0.4 horizontal dashed line. However, in the bottom left corner in the right panel of Fig. \ref{fig:Amod_offset_visual_class1}, two galaxies (31 and 45) that are visually classified as galaxies with a disturbed or an unsettled \hi disc nevertheless have a low \amod value. These galaxies have a low \amod value because \amod is calculated within the green contour, which corresponds to 25 $\times$ 10$^{19}$ cm$^{-2}$, which is above the 3$\sigma$ level (because it is inside the outermost black contour). These galaxies, however, have disturbed / unsettled \hi discs when we consider the outer \hi contour, which corresponds to the 3$\sigma$ column density. This illustrates that the choice of 25 $\times$ 10$^{19}$ cm$^{-2}$ means that for some galaxies, the visual classification based on the outer \hi disc will not correspond to the \amod value calculated with pixels above the 25 $\times$ 10$^{19}$ cm$^{-2}$ contour. Furthermore, in the left panel of Fig. \ref{fig:Amod_offset_visual_class1} lie three galaxies with \amod $>$0.4, but they are visually classified to have settled \hi discs. For galaxies 147 and 120, the \hi disc seems somewhat offset from the optical centre. An \hi cloud is detached from the \hi disc in galaxy 191, but it was included in the calculation of \amod. However, the visual impression of the \hi discs of these galaxies suggests that they are settled. This illustrates that the \amod values and the visual classifications both are valuable tracers of \hi morphologies and largely follow each other.

\begin{figure}[t!]
\centering
 {
\includegraphics[width=0.5\textwidth]{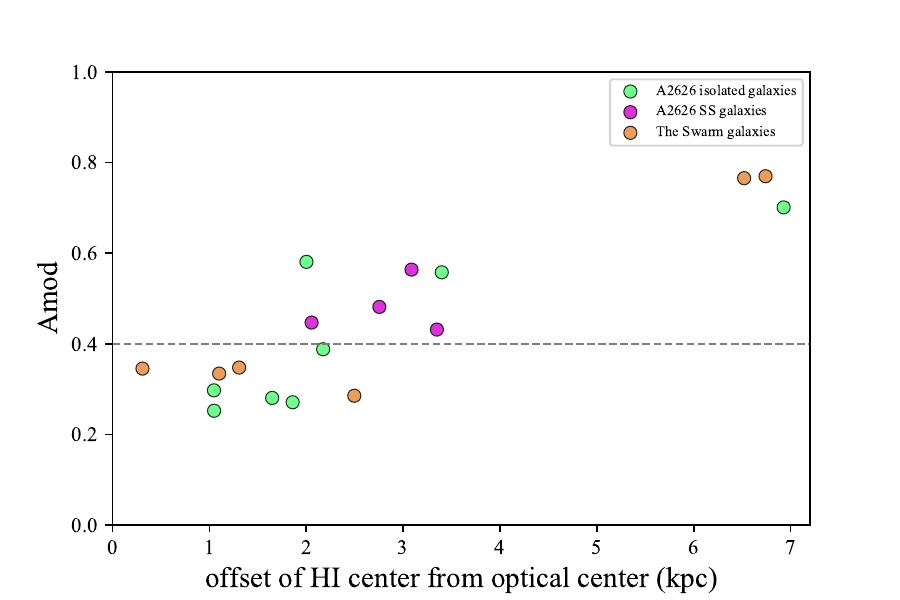} 
  } 
  \caption{Distribution of \amod and \hi offset, colour-coded with three different environments for the galaxies residing in and around A2626. The green, magenta, and orange circles represent non-substructure and substructure galaxies in A2626, and galaxies in the Swarm, respectively. Substructure galaxies seem to have more asymmetric \hi discs with high \amod compared to non-substructure galaxies in A2626.} 
  \label{fig:Amod_offset_diff_env} 
 \end{figure}
 
 As a next step, we considered the environment as a parameter in the same \amod versus \hi offset plot. We used three different colours to represent the three different environments, as described in Sec. \ref{subsec:diff_envs}. In Fig. \ref{fig:Amod_offset_diff_env}, the green, magenta and the orange circles represent the non-substructure galaxies in A2626, the substructure galaxies in A2626, and the Swarm galaxies, respectively. Although the statistics is limited, all the galaxies in the substructure for which we have reliable \amod values are \hi asymmetric galaxies. At the same time, only a fraction of the non-substructure galaxies in A2626 have an \hi asymmetric disc. The occurrence of more \hi asymmetric galaxies in substructures of A2626 (all magenta symbols are above the dashed line) might be due to more effective tidal interactions in the substructure environment than the non-substructure galaxies in A2626 experience. However, galaxies in the Swarm and non-substructure galaxies in A2626 are found to have both disturbed and undisturbed \hi discs.


\section{\hi\ deficiency in A2626 and the Swarm galaxies}
 \label{sec:HI_def}

In addition to exploring different \hi morphological classifications and their significance, we explored the \hi deficiency as another indicator of environmentally induced gas depletion and removal processes. We examined the \hi deficiency of a galaxy as a function of its projected distance from the centre of A2626, its position in the phase-space diagram of A2626, and its \hi morphology.

\hi deficiency is defined as a logarithmic quantity (Haynes \& Giovanelli 1983), a difference between the log of the expected \hi mass and the log of the observed \hi mass of a galaxy,

\begin{equation}
    HIdef = log[M_{HI}^{exp}] - log[M_{HI}^{obs}]
.\end{equation}
 
 \begin{figure*}[ht!]
 {
    \includegraphics[width=\textwidth]{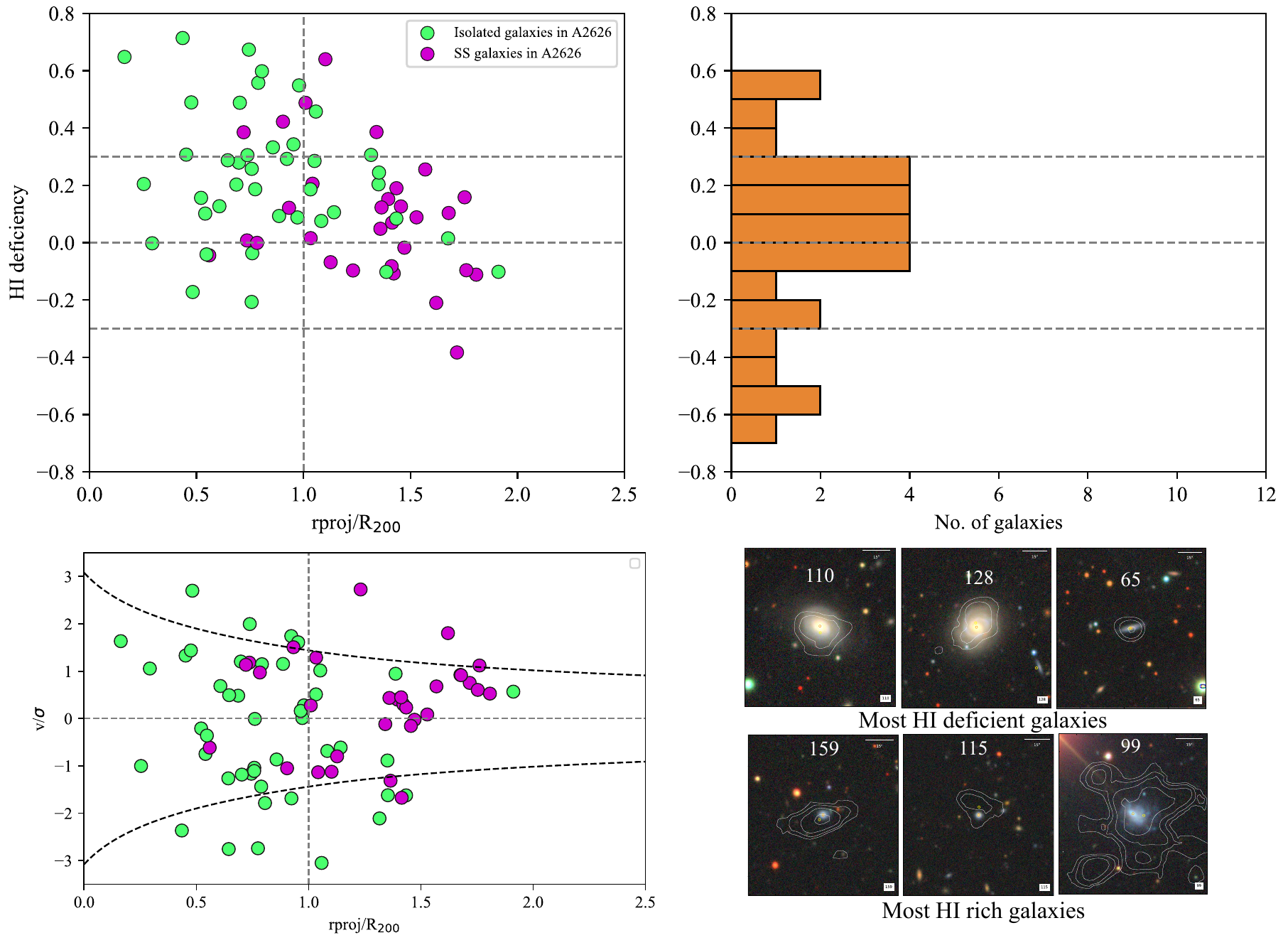} 
  } 
  \caption{\footnotesize{Gas properties, velocities and morphologies of galaxies in A2626 and the Swarm.  Top left: \hi deficiency vs projected distance normalised by R200 for the non-substructure (green) and substructure (magenta) galaxies in A2626.  The horizontal dashed lines present the range of \hi deficiencies of the field galaxies.  Top right: Histogram of the distribution of \hi deficiencies of galaxies in the Swarm.  Bottom left: Distribution of the non-substructure (green) and substructure (magenta) galaxies in A2626 in projected phase-space.  The dashed black
 lines indicate the escape velocity.  Bottom right: Six \hi maps overlaid on DECaLS colour images for the most \hi-deficient (top three) and the most \hi-rich (bottom three) galaxies.  The \hi-deficient galaxies seem to be bright and yellowish with an offset or truncated \hi discs.  The \hi rich galaxies are fainter and bluish, with extended \hi discs.
} }
  \label{fig:HIdef_plot1} 
 \end{figure*}

 \noindent Thus \hi deficiency is positive for \hi -deficient galaxies and negative for galaxies with excess \hi gas. $M_{HI}^{exp}$ is generally calculated from \hi-optical scaling relations (e.g. \citealt{Haynes1984, chamaraux1986, Batuski1985, solanes1996, denes2014}). We used the scaling relation from \citet{denes2014} (see their Table 3), a multi-wavelength scaling relation between the \hi content and the optical diameter of the galaxies. The scaling relation of \citet{denes2014} is based on the SDSS r-band relation, which uses the Petrosian radius, 
 
 \begin{equation}
   log[M_{HI}^{exp}] = \alpha_{\lambda} + \beta_{\lambda} log[D_{\lambda}]  
 ,\end{equation}
 
\noindent where $M_{HI}^{exp}$ is the expected \hi mass, $D_{\lambda}$ is the optical diameter (in kpc) in a particular band, and $\alpha_{\lambda}$ and $\beta_{\lambda}$ are the parameters for this band. To calculate the expected \hi mass, we adopted the diameters (D$_{r}$=2R$_{25}$) provided by J. Healy (private communication) as determined from the DECaLS r-band images, where R$_{25}$ is the radius of the 25th mag/arcsec$^{2}$ isophote.

\subsection{\hi deficiency in substructures}
The top left panel of Fig. \ref{fig:HIdef_plot1} shows the \hi deficiency of non-substructure and substructure galaxies vs. their projected distance from the centre of A2626, normalised by R$_{200}$ of A2626. R$_{200}$ is the projected radius of a sphere with a mean density 200 times the critical density of the Universe. The \hi deficiencies of the Swarm galaxies are presented in the orange histogram in the top right panel of Fig. \ref{fig:HIdef_plot1}. First of all, we observe a correlation between \hi deficiency and the projected distance from the centre of A2626 (Spearman's coefficient= -0.377, p-value= 0.0012). This clearly illustrates that galaxies become \hi deficient towards the cluster core. Moreover, within 1.5 R$_{200}$, the most \hi -deficient galaxies seem to be the non-substructure galaxies. On average, non-substructure galaxies are more \hi deficient than sub-structure galaxies in A2626 ($< HIdef>_{non-ss}$ = 0.27, $<HIdef>_{ss}$=0.15). We note, however, that the substructure galaxies are more prevalent at larger clustercentric radii. The six panels in the bottom right corner of Fig. \ref{fig:HIdef_plot1} illustrate \hi maps overlaid on DECaLS colour images for the most \hi -deficient (top three) and the most \hi -rich (bottom three) galaxies. Focusing on these, we observe that the \hi -deficient galaxies are bright and have a yellowish colour, and they have small or truncated \hi discs. In contrast, the \hi -rich galaxies are fainter, with a bluish colour, and have extended \hi discs. When we restrict the sample to the cluster galaxies (non-substructure plus substructure), the Pearson correlation coefficient between the g-r colour and \hi deficiency is 0.30, with a p-value of about 0.01, which means that there is a weak but statistically significant (> 95\% confidence level) correlation between \hi deficiency and colour for cluster galaxies. This quantitatively supports our visual impression that the most \hi-deficient galaxies are typically redder than the most \hi-rich ones.

\begin{figure}[t!]
\begin{center}

 {
    \includegraphics[width=0.45\textwidth]{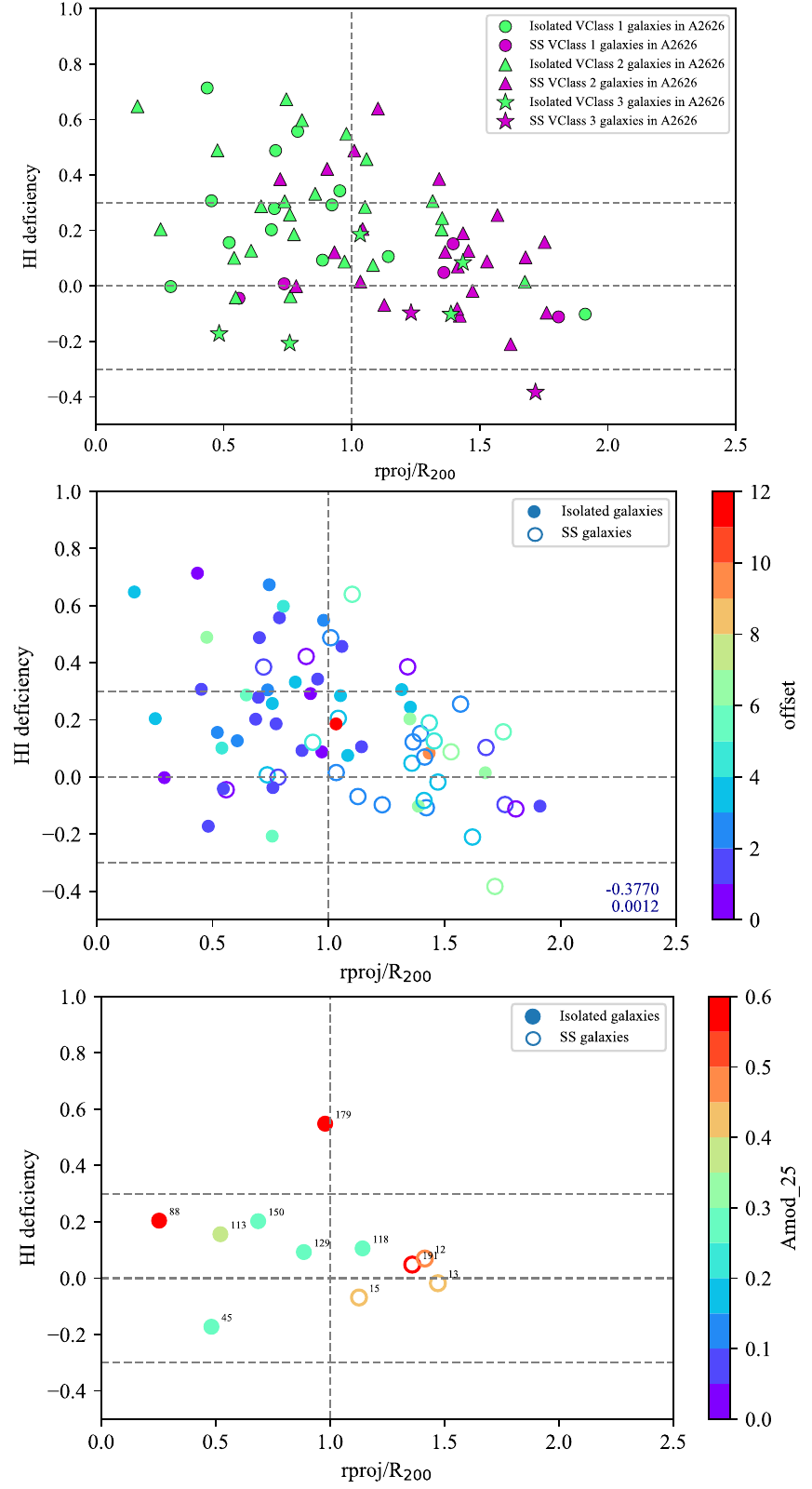} 
  } 
  \caption{\footnotesize{\hi deficiency vs projected distance normalised by R$_{200}$ (similar to Fig. \ref{fig:HIdef_plot1}), with symbols and colour codes to include additional information regarding \hi morphologies of galaxies in A2626. Top panel: Different symbols (shown in the top right corner) represent non-substructure (green) and substructure (magenta) galaxies of different visual classes. Middle panel: Colour scale (colour bar shown on the right) representing \hi offsets of non-substructure (filled circles) and substructure (open circles) galaxies. Bottom panel: Colour scale (colour bar shown on the right) representing the \amod value of non-substructure (filled circles) and substructure (open circles) galaxies.}} 
  \label{fig:HIdef_plot2} 
  \end{center}
 \end{figure}

\subsection{\hi deficiency in phase-space}

In the bottom left panel of Fig. \ref{fig:HIdef_plot1}, we show the phase-space diagram for non-substructure and sub-structure galaxies in A2626. The dashed black lines indicate the escape velocity for A2626 galaxies, calculated using the formalism from \citet{Jaffe2015}. We assumed a concentration index C = 6, and the mass M$_{200}$ enclosed by R$_{200}$ was calculated using

\begin{equation}
    M_{200} = \frac{4}{3} \pi R^{3}_{200}200\rho_{c}
\end{equation}

\noindent where $\rho_{c}$ is the critical density. The vertical dashed grey line in the phase-space diagram represents R$_{200}$ of A2626. In the phase-space diagram, there is no clear additional trend in the distribution of non-substructure and substructure galaxies, except that
 substructure galaxies are not found near the core of the cluster and tend to have positive velocity offsets in the cluster outskirts.
 
\subsection{\hi deficiency and \hi morphology}
In this subsection we explore possible relations between the \hi deficiencies and the \hi morphologies of galaxies in order to gain insights in the physical mechanisms that may be responsible for the \hi deficiencies. As a next step, in Fig. \ref{fig:HIdef_plot2}, we explore the same \hi deficiency vs projected distance plot with colour codes and symbols to include additional information regarding the \hi morphologies of these 76 galaxies. 
\newline
\textbf{\hi deficiency and visual classifications}

\noindent In the top  panel of Fig.  \ref{fig:HIdef_plot2}, we present the \hi -detected galaxies of the three visual classes as defined in Sec. \ref{subsec:visual_classes} with circles, triangles, and stars. The green and magenta symbols in each class stand for the non-substructure and sub-structure galaxies in the A2626 cluster. Interestingly, the Vclass 3 or unsettled galaxies (starss) seem to have \hi deficiencies similar to the field galaxies (i.e. within the three horizontal dashed lines). Moreover, substructures contain more disturbed / unsettled galaxies (triangles and stars) than the non-substructure galaxies in A2626, as we showed in Sec. \ref{subsec:visual_classes}. However, there is no obvious correlation of the visual classes of the \hi detected galaxies with their \hi deficiencies or with their projected distance from the cluster centre. Thus, the visually classified \hi morphologies are not directly related to the \hi deficiencies of the galaxies or their location in the A2626 cluster. 
\newline \textbf{\hi deficiency and \hi offsets}

\noindent In the middle panel of Fig. \ref{fig:HIdef_plot2}, we again plot the \hi deficiency vs projected distance from the centre of A2626 for the \hi -detected galaxies. The colour-coding follows the offset (in units of kiloparsec) of the \hi centres of these galaxies with respect to the optical centres. The filled and open circles are non-substructure and substructure galaxies, respectively. There is no correlation between \hi deficiency and \hi offset for the galaxies in A2626. This implies that the \hi -deficient galaxies do not necessarily have \hi offset discs and vice versa.
\newline \textbf{\hi deficiency and \amod}

\noindent In the bottom panel of Fig. \ref{fig:HIdef_plot2}, we examine the \hi deficiency vs projected distance. The colour-coding indicates the \amod values of the \hi discs of galaxies in A2626. Because \amod is meaningful for only 11 well-resolved galaxies in A2626 with high signal-to-noise ratio, we plot only those galaxies. The filled and open symbols are non-substructure and sub-structure galaxies in A2626. Similar to the two other panels in Fig. \ref{fig:HIdef_plot2}, there is no clear trend between \hi deficiency, projected distance from the cluster centre, and \amod in the galaxies in A2626. Interestingly, the most \hi -deficient galaxy in this plot is also the most \hi -asymmetric galaxy with a high \amod value (\amod = 0.6).

\bigskip Considering all panels of Fig. \ref{fig:HIdef_plot2}, we conclude that there are no further obvious correlations between \hi deficiency, \hi morphology, and the location of the galaxy in the cluster except for
the clear trend between \hi deficiency and the projected distance from the
cluster centre, except that Vclass 3 galaxies are not \hi deficient.


\section{Star formation rates and depletion times in A2626 and the Swarm galaxies}
\label{sec:sfr_dep_time}

\begin{figure*}[t!]
 {
    \includegraphics[width=\textwidth]{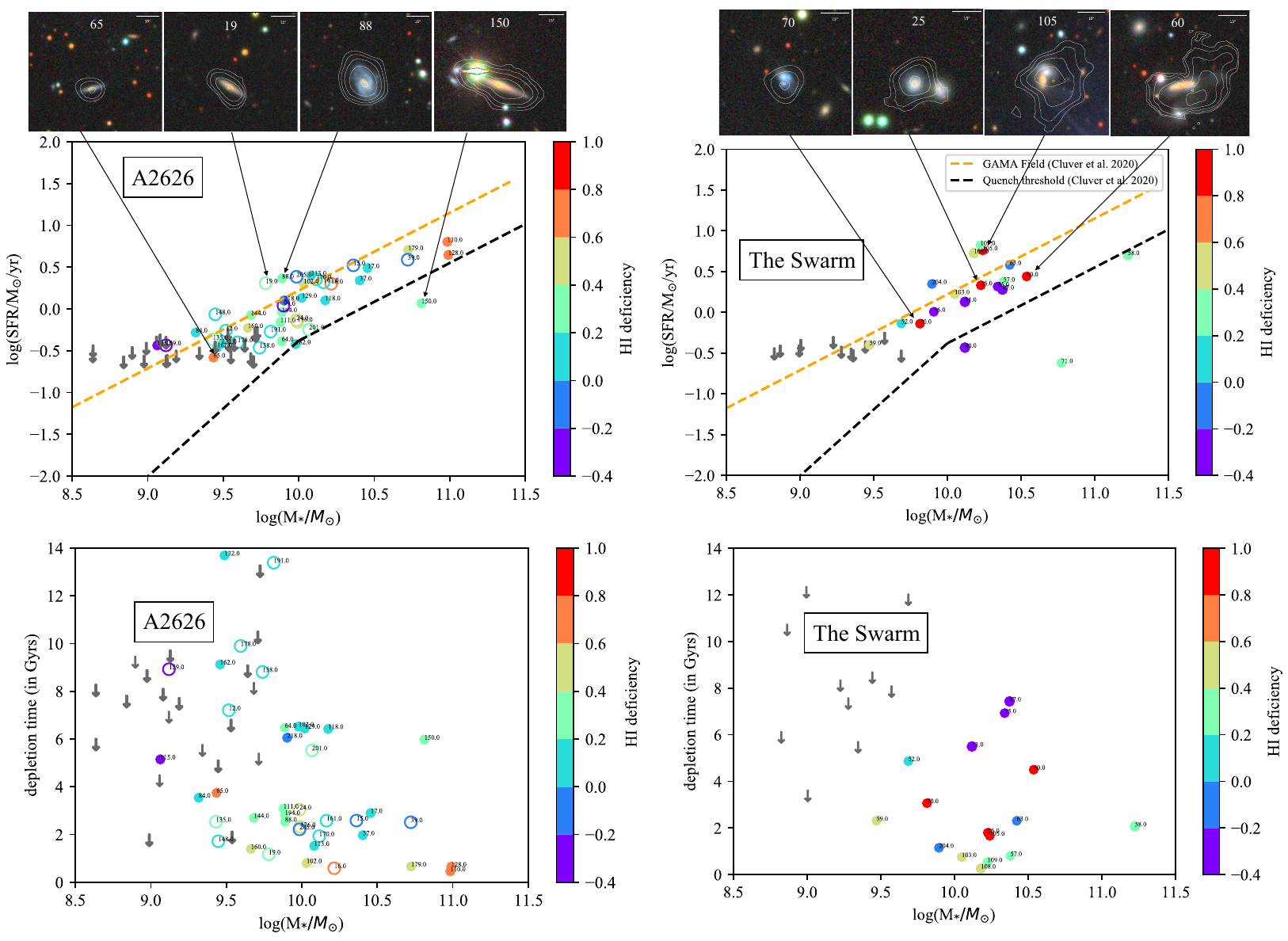} 
  } 
  \caption{\footnotesize{Star formation rates and gas depletion times as a function of stellar mass and \hi deficiency for galaxies in A2626 and the Swarm. Top left panel: SFMS for the non-substructure (filled circles) and substructure (open circles) galaxies in A2626. The SFMS relation and quenching threshold are taken from \cite{cluver2020}, which were calibrated using the WISE data and with the same methods as were used for the stellar mass and SFR calculations for the galaxies in our survey. The colour scale (shown on the right) represents the \hi deficiency for the galaxies. The downward arrows are 2$\sigma$ upper limits on the SFR from WISE observations. In the top panels, we show \hi maps overlaid on DECaLS colour images for some outlier galaxies. Top right panel: Similar to the plot on the top left, but for galaxies in the Swarm. Bottom left panel: \hi depletion time vs stellar mass of the non-substructure (filled circles) and substructure (open circles) galaxies in A2626. The colour scale presents the \hi deficiency of the galaxies. Bottom right panel: Similar to the bottom left plot, but for the galaxies in the Swarm.} }
  \label{fig:SFR_Mstar_tdep_HIdef} 
 \end{figure*}

\subsection{Galaxies on the SFMS }
 
We combined the information on star formation rates (SFR) and stellar masses with the \hi properties of the galaxies in A2626 and the Swarm to investigate the extent of the effect of different environments on the star formation properties of the \hi -detected galaxies. The star formation activity and the stellar mass of a galaxy are related in a systematic way that is described by a well-established scaling relation, the star formation main sequence (SFMS; e.g. \citealt{brinchmann2004, Noeske2007, elbaz2007, speagle2014, tomczak2016}). We investigated the location of the galaxies in different environments (non-substructure and sub-structure galaxies in A2626, and the Swarm galaxies) with respect to the SFMS. We adopted the SFMS and the star formation quenching threshold relation for non-group galaxies from \citet{cluver2020}, which were calibrated using WISE data. Thus, any trend we observe for the group and cluster galaxies might be due to the impact of the group or cluster environments on the SFR as compared to the galaxies that are not a part of any group (i.e. the galaxies in the \citet{cluver2014} relation). For consistency, we also adopted stellar masses and SFRs based on WISE data derived with the same method as developed by \citet{cluver2014} (see Sect. 1.3).

We also investigated the relation between \hi depletion time and stellar mass for galaxies in different environments as a function of \hi deficiency. The depletion time is the time required for a galaxy to deplete its \hi gas due to star formation and is defined as $t_{dep}$ = M$_{HI}$/ SFR. The depletion timescale of normal star-forming galaxies spans a range of $\sim $2-10 Gyr \citep{kennicutt1989, Kennicutt1998, bigiel2008}.

\subsection{HI deficiency versus SF deficiency }

 In Fig. \ref{fig:SFR_Mstar_tdep_HIdef} we plot the SFMS in the top two panels and the depletion time versus stellar mass in the bottom two panels for the galaxies in A2626 and the Swarm. To represent the two different environments in A2626, we used filled and open circles for the non-substructure and substructure galaxies in A2626 in the top and bottom left panel of Fig. \ref{fig:SFR_Mstar_tdep_HIdef}, respectively. In all the panels, we colour-code the \hi deficiencies. The dotted orange and black lines are the SFMS and the quenching threshold from \citet{cluver2020}, respectively. The downward-pointing arrows are the 2$\sigma$ upper limits on the SFR. Interestingly, galaxies in all three environments, the non-substructure and substructure galaxies in A2626, and the Swarm galaxies, are mostly below the SFMS from \citet{cluver2020}, and thus show an overall SF deficiency.
 
 \begin{figure}
 {\includegraphics[width=0.45\textwidth]{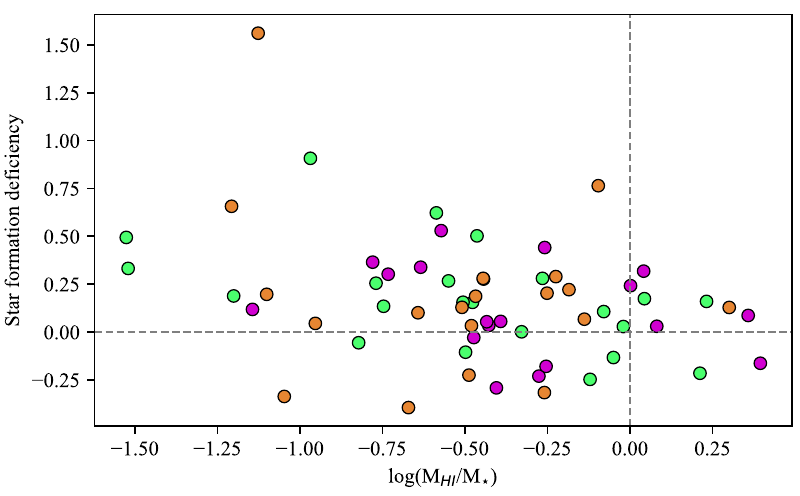} 
  } 
  \caption{SF deficiency in A2626 and the Swarm galaxies. Galaxies above the horizontal dashed line lie below the SFMS and can therefore be considered to have a positive SF deficiency. The colour-coding is similar to the colour-coding in Fig. \ref{fig:Amod_offset_diff_env}.}
  \label{fig:offset_SFMS_A2626_Swarm} 
 \end{figure}
 
  \begin{figure*}[ht!]
{\includegraphics[width=\textwidth]{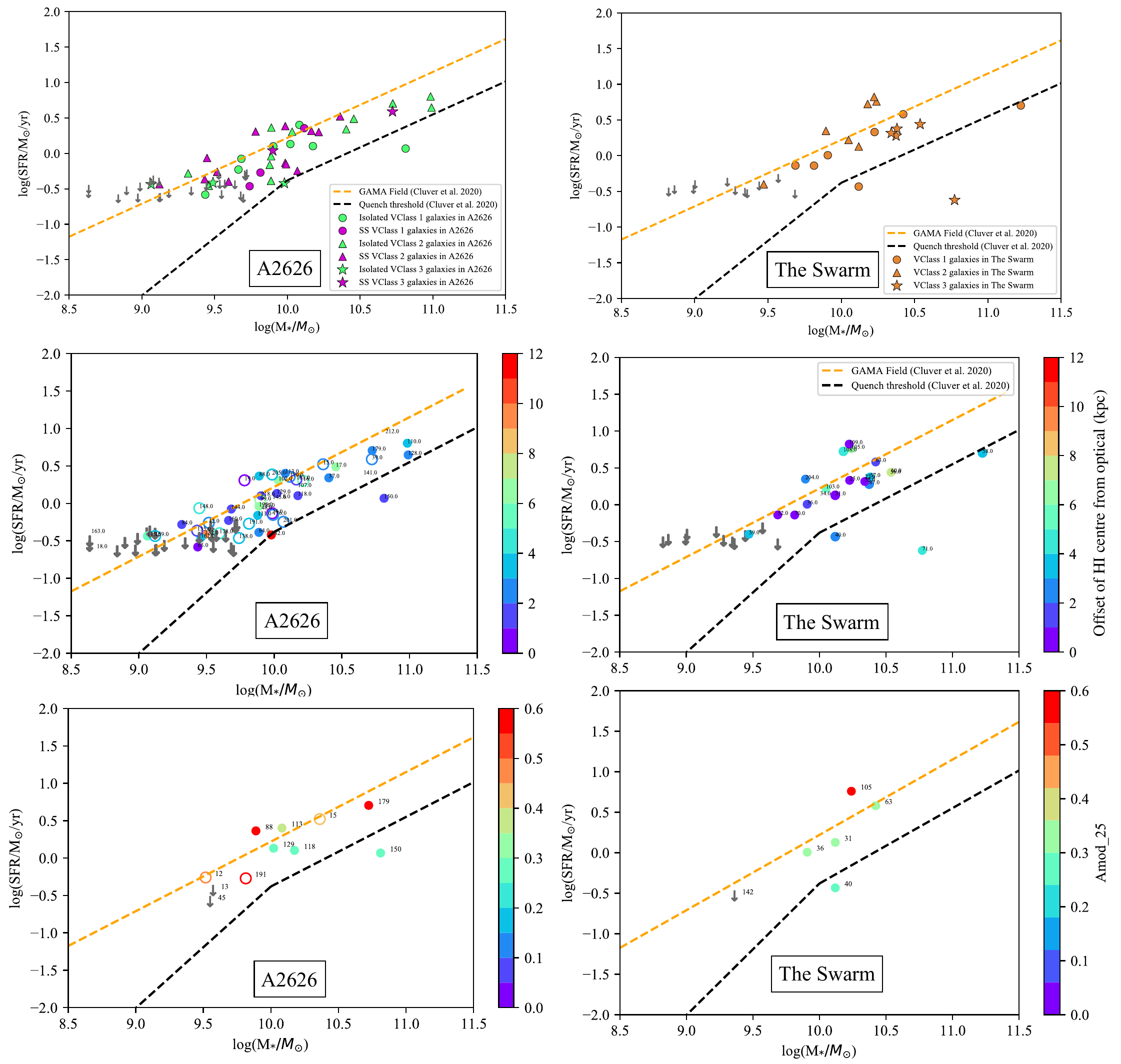}} 
  \caption{Similar to Fig. \ref{fig:SFR_Mstar_tdep_HIdef}, with additional information regarding the \hi morphologies of galaxies in A2626 and the Swarm.} 
  \label{fig:SFR_Mstar_tdep_Amod} 
 \end{figure*}
 
 To determine whether there is any difference in the distribution of galaxies in the three environments with respect to the SFMS, we plot in Fig. \ref{fig:offset_SFMS_A2626_Swarm} the SF deficiency (defined as the SFMS minus the measured SF of a galaxy) in the three environments (green: non-substructure galaxies in A2626, magenta: substructure galaxies in A2626, and orange: the Swarm galaxies). A positive SF deficiency (galaxies above the horizontal dotted line) indicates that these galaxies are below the SFMS, and a negative SF deficiency (galaxies below the horizontal dotted line) indicates that these galaxies are above the SFMS. Along the horizontal axis, we plot M$_{HI}$/M$_{\star}$ to determin whether the SF deficiency relates to the relative \hi content of these galaxies. We do not observe any differences for the galaxies in different environments. This means that the star formation activity of galaxies with a similar stellar mass does not depend on the environment. Galaxies with a low relative \hi content tend to be SF deficient, which is expected. 
 
 \subsection{\hi depletion time versus \hi morphology}
 In the bottom panels of Fig. \ref{fig:SFR_Mstar_tdep_HIdef}, no clear trend between the depletion time and stellar mass for the galaxies is visible, regardless of their environment. The downward-pointing arrows are 2$\sigma$ upper limits on the depletion times. The figures show that no galaxies have high stellar masses and long depletion times. Massive galaxies tend to have a high SFR, as expected from the SFMS, and thus exhaust their \hi fuel relatively fast. The depletion time is weakly correleated with \hi deficiency. However, we note that the most \hi -deficient galaxies in A2626 have a short depletion time, while the three galaxies in the Swarm with the longest depletion times (purple symbols) have a negative \hi deficiency. 
 
 \subsection{\hi morphology versus SF deficiency}
 
 In Fig. \ref{fig:SFR_Mstar_tdep_Amod} we present similar SFMS plots in six different panels for three different environments, using the same symbols as Fig. \ref{fig:SFR_Mstar_tdep_HIdef}. The only differences are that in the two top panels of Fig. \ref{fig:SFR_Mstar_tdep_Amod}, we use different symbols for the three visual classifications. In the middle panels, we colour-code with \hi offset, and in the bottom panels, we colour-code with \amod. Here again, we do not observe any meaningful correlation between the SF deficiency and \hi offset or \amod as a function of environment or stellar mass.

\section{Interesting galaxies}
\label{sec:interesting_gals}

Before we discuss our results in detail, we first present the detailed \hi maps of several interesting galaxies to illustrate the diversity of the \hi morphologies. These galaxies are resolved in our \mk \hi observations, and we compare them with the existing DECaLS optical colour images, their star formation rates from WISE observations, and their location in phase-space to infer the ongoing physical mechanisms acting on these galaxies. 

\begin{figure*}[t!]
 {
    \includegraphics[width=0.9\textwidth]{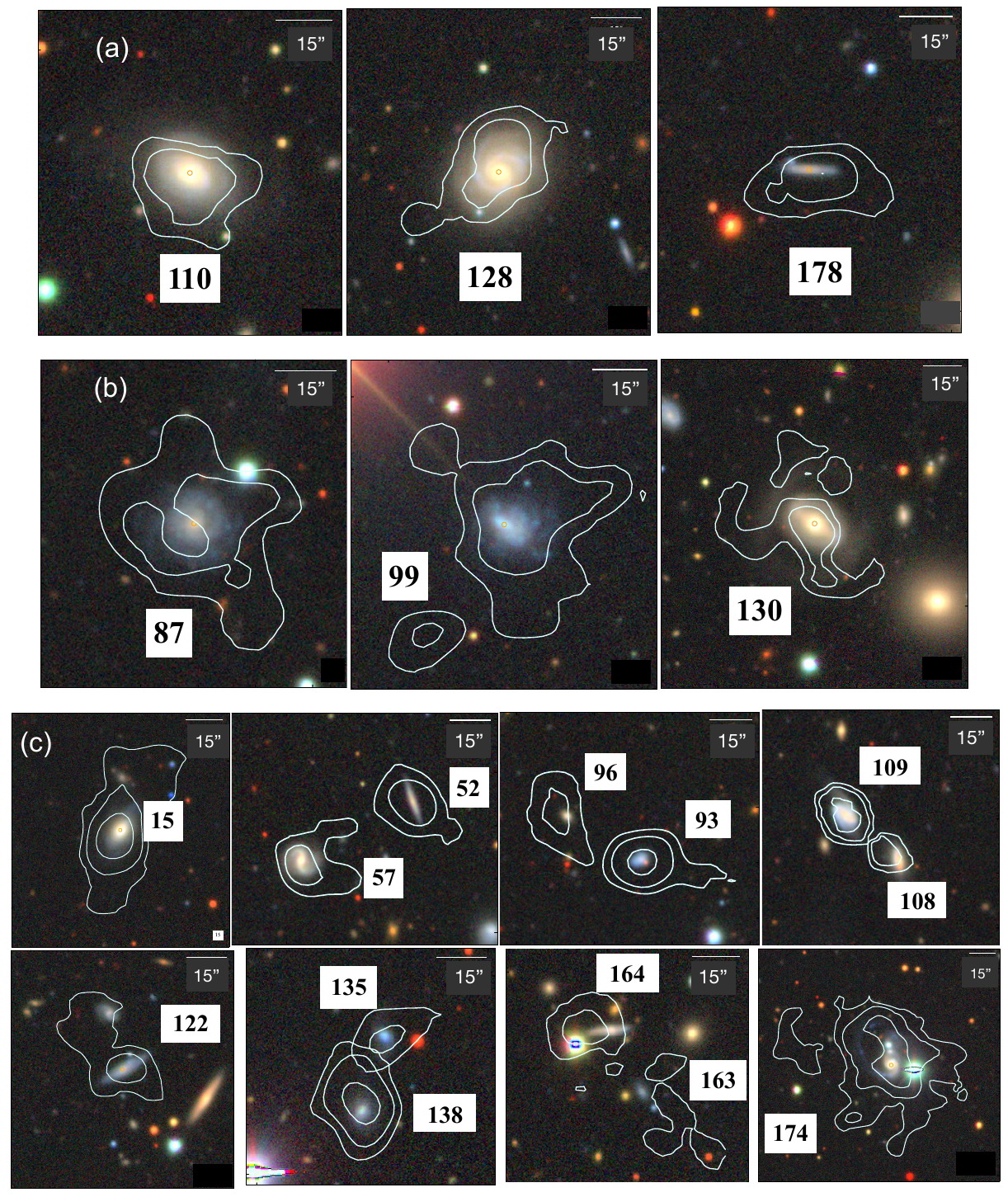} 
  } 
  \caption{Interesting galaxies with diverse \hi morphologies in and around A2626.} 
  \label{fig:interesting_gals} 
 \end{figure*}

 \subsection{Potential ram-pressure-stripped galaxies}
 
 In Fig. \ref{fig:interesting_gals} (a) we present several galaxies that might be experiencing ram-pressure stripping or other effects such as thermal evaporation or starvation. Galaxies 110 and 128 are non-substructure galaxies in A2626, are located close to the cluster core (within R$_{200}$), and are fully exposed to the ICM. Moreover, they are highly \hi deficient and have small \hi discs with yellowish optical discs indicating a quenched star formation, which is confirmed by their low star formation rates inferred from the WISE fluxes. Hence, they probably have experienced  ram-pressure stripping or thermal evaporation for a long time, which has resulted in the removal of a significant amount of \hi gas. Galaxy 178 is located in a substructure in A2626. Although it is not very close to the cluster core ($\sim$ 1.5 R$_{200}$), the \hi morphology indicates ongoing ram-pressure stripping. Its \hi contours are compressed on the northern side of the \hi disc, and the outer \hi contour seems to suggest that the \hi gas is pushed to the south. Interestingly, galaxy 178 is moving at a velocity similar to the mean velocity of the cluster and has a low star formation rate. Therefore, we may conclude that galaxy 178 is a backsplash galaxy, but the \hi content of the galaxy suggests that it is still on its first-infall trajectory.
 
 \subsection{Galaxies with unsettled \hi discs}
 
 In Fig. \ref{fig:interesting_gals} (b) we present some well-resolved galaxies with unsettled \hi discs (Vclass 3) and high \amod values (\amod$>$0.5). Galaxy 87 is a face-on system in the Swarm with a regular optical morphology, but an extended unsettled \hi disc. We note a nearby companion $\sim$ 70" to the south-west. Galaxy 99 is in a substructure in A2626. Not only does it have an unsettled \hi disc, but it is also optically highly disturbed. It is likely that it has experienced tidal interaction with another galaxy in the substructure. Galaxy 130 is a foreground galaxy at a redshift of z$_{HI}$=0.0382, and therefore is excluded from our overall analysis. Its optical morphology is that of an early-type barred ring galaxy. \hi emission is not only found in the bright central regions of the galaxy, but clumps of \hi gas are also found in the faint outer stellar ring surrounding the bar, similar to NGC 4736 \citep{bosma1977}.
 
\subsection{Interacting galaxies}
In Fig. \ref{fig:interesting_gals} (c) we present examples of interacting galaxies in our sample that are confirmed to have similar velocities based on the MMT as well as \hi redshifts \citep{HealySS2021}. Among the interacting galaxies shown in Fig. \ref{fig:interesting_gals} (c), galaxies 15, 122, 135, 138, 163 and 164 are in A2626 itself. Galaxies 52 and 57 are in the Swarm, while galaxies 93, 96, 108, 109, and 174 are in the background cluster A2637. The \hi emission of galaxies 15, 122, and 174 is blended with that of their companion galaxies. We conclude that the gravitational interactions are prevalent in both the group and the cluster environment in and around A2626.

\section{Discussion}
\label{sec:discusiions}

\cite{HD2021} presented the first results of our \mk observations of A2626, including online atlas pages of the \hi detected galaxies. In this work, we characterised \hi asymmetries in various ways and calculated \hi deficiencies. We then related these to local and global environment and specific galaxy properties such as SFR and stellar mass. As discussed in the review paper by \cite{cortese2021}, during the last few billion years in what today are passive galaxies, it is inevitable that most of the primary quenching mechanisms must have played a role in shaping their properties (e.g. \citealt{mcgee2009, han2018}). Because multiple processes can be at play simultaneously, it is difficult to identify the dominant process observationally. Still, as mentioned before, by exploring various gas and star formation properties of galaxies, we can investigate the plausible physical mechanisms that act on their \hi gas reservoirs. Mechanical gas-removal processes such as ram-pressure stripping and tidal stripping are likely to push or pull the \hi gas away from the stellar body, causing offset and asymmetric \hi discs or unidirectional tails. Sometimes, however, when they are exposed to ram-pressure for a long time, \hi discs lose most of the gas from the outer disc and become truncated. Processes such as starvation, harassment, or thermal evaporation mainly occur in dense cluster environments and also give rise to truncated or small \hi discs, rather than offset \hi discs. If the \hi -deficient galaxies have settled small \hi discs, a gas-removal process such as ram-pressure stripping therefore has probably occurred a while ago and / or other more subtle processes may be at play, such as starvation or thermal evaporation. In this section, we address some of these observational findings and the questions raised by our analysis. 

\subsection{Trends with projected distance}
First of all, the clearest result we found while examining the \hi properties of the galaxies in A2626 is the increase in \hi -deficient galaxies towards the cluster core in Fig. \ref{fig:HIdef_plot1}. This trend probably indicates an increase in gas removal efficiencies in the cluster core, similar to what previous observations have found (e.g. \citealt{Solanes2001, chung2009}). Moreover, \hi -detected galaxies in the substructures are mainly found in the cluster outskirts. Interestingly, beyond R$_{200}$, both non-substructure and substructure galaxies in A2626, as well as the galaxies in the Swarm, have a similar range of \hi deficiencies. This indicates pre-processing in some of the substructure galaxies in A2626 and in the Swarm galaxies before they are fully exposed to the intra-cluster medium. 

 Next, we investigated whether the shapes of the outer \hi discs are also affected by the different physical mechanisms driving \hi deficiencies in the various environments in and around A2626. In Fig. \ref{fig:HIdef_plot2} we explored how \hi deficiency versus projected distance from the cluster centre depends on visual classes (top panel), \hi offset (middle panel), and \amod (bottom panel). In the top panel of Fig. \ref{fig:HIdef_plot2}, the Vclass 3 or unsettled galaxies (stars) seem to have \hi deficiencies similar to field galaxies and even tend to be slightly more gas rich on average. Thus, these unsettled galaxies are still in an active stage of evolution, and their \hi discs still contain a large gas reservoir for star formation. The occurrence of more \hi asymmetric galaxies in the substructures of A2626 might be due to more effective tidal interactions among the neighbouring galaxies in the substructure environment compared to the non-substructure galaxies in A2626. These interactions, however, do not remove the \hi gas efficiently, making the galaxies in the substructures only moderately \hi deficient. This indicates ongoing pre-processing in the substructure galaxies in A2626 before they virialise in the cluster core. Considering the most \hi -deficient galaxies (HIdef$>$0.3) within $\sim$ 1.5 R$_{200}$, we find that 12 out of the 17 galaxies have disturbed \hi discs (Vclass 2), indicating that they may be still under the influence of ram-pressure stripping. The other 5 galaxies have a settled \hi discs (VClass 1), but their \hi discs seem to be small with respect to their stellar discs, indicating that they are in an  advanced stage of ram-pressure stripping or under the influence of processes such as starvation or thermal evaporation. In the middle panel of Fig. \ref{fig:HIdef_plot2}, we do not observe any clear correlation between \hi deficiency and the amount of \hi offset. This can be understood by realising that the direction of ram-pressure stripping is not always perpendicular to the line of sight, and the small offsets induced by harassment would not have a preferred direction either. In the bottom panel of Fig. \ref{fig:HIdef_plot2}, we find no clear relation of \hi deficiency with \amod , although the most \hi -deficient galaxy is highly asymmetric.
 
\subsection{Trends with respect to the SFMS}
 In Figs. \ref{fig:SFR_Mstar_tdep_HIdef} and \ref{fig:SFR_Mstar_tdep_Amod} we explored the location of the A2626 and the Swarm galaxies with respect to the SFMS, as well as the relation between their \hi depletion times and stellar masses. That the galaxies in all three environments (both the substructure and non-substructure galaxies in A2626 and the Swarm galaxies) tend to be below the SFMS from \cite{cluver2020} means that both the cluster and group environments induce a slightly lower SFR than the SFR expected for normal galaxies. There is no clear relation between \hi deficiency and offset from the SFMS, that is, SF deficiency, for either the cluster or the Swarm galaxies. This implies that the mechanisms causing \hi deficiencies probably do not immediately impact the SFR and vice versa, regardless of the environment. 
 
 However, some galaxies (e.g. substructure galaxy 19 and non-substructure galaxy 88 in A2626 in Fig. \ref{fig:SFR_Mstar_tdep_HIdef}) have a higher SFR than the SFMS. Their \hi morphologies appear to be fairly regular (insets of the top left panel in Fig. \ref{fig:SFR_Mstar_tdep_HIdef}), while they are moderately \hi deficient and have short \hi depletion times. Galaxy 88 was identified as a candidate jellyfish galaxy from the B-band optical image by \cite{Poggianti2016}. However, because its \hi morphology is regular and it is only moderately \hi deficient, this galaxy was later identified as a non-jellyfish galaxy in \cite{Deb2022}. We further inspected \hi maps of some other outlier galaxies from the SFMS. Galaxy 150 is below the quenching threshold, but it is not particularly \hi deficient. The \hi disc of this galaxy probably has a low column density that is below the star formation threshold, which results in a long depletion time because the SFR is lower.
 
 We also investigated the \hi maps of the most \hi -deficient galaxies in A2626 (galaxies 16, 65, 110, and 128; see Figs. \ref{fig:HIdef_plot1} and \ref{fig:SFR_Mstar_tdep_HIdef}). All four are located below the SFMS. Although the \hi disc of galaxy 65 seems small and regular, the other three galaxies have asymmetric \hi morphologies. As already shown with \hi contours overlaid on the optical colour images in Fig. \ref{fig:HIdef_plot1}, all these galaxies have small, offset, or truncated \hi discs and yellowish optical colours. As mentioned before, these three galaxies are below the SFMS, signifying low relative SFRs, although they are still above the star formation quenching threshold. However, because of their high \hi deficiencies, they have very short depletion times. Thus, all these four galaxies probably are in the last stages of exhausting their gas and are on their way to drop below the star formation quenching threshold. 
 
 Focusing on the \hi morphologies of the most \hi -deficient galaxies in the Swarm (galaxies 25, 60, 70, and 105 in Fig. \ref{fig:SFR_Mstar_tdep_HIdef}), we note that all of them have nearby companions, and two of these galaxies (galaxies 60 and 105) have asymmetric or offset \hi discs. From the optical colour, galaxy 105 seems dusty, which might cause some internal reddening. Galaxy 60 is tidally interacting, with a visible stellar stream to the west, similar to the tidally interacting system Arp 282 \citep{Zaragoza-Cardie2018}. Although we lack optical or \hi redshift information for the companions of galaxies 25 and 70, these galaxies clearly experience tidal interactions given their \hi morphologies.

Finally, we investigated the relation between SF deficiency and \hi morphology using the three different morphological definitions (visual classes, \hi offset, and \amod; see Fig. \ref{fig:SFR_Mstar_tdep_Amod}). That there is no correlation between SF deficiency and \hi offset or \amod implies that an enhancement or quenching of their star formation rates are not directly related to their \hi morphologies or to the environment in which they reside.  
 
\bigskip We conclude this discussion by noting that the \hi content and \hi morphologies of galaxies are useful tracers of various environmental effects. Although the galaxies are \hi deficient close to the cluster centre, these \hi -deficient galaxies do not always have disturbed or unsettled \hi discs. Consequently, in addition to processes such as ram-pressure stripping and tidal interactions, more subtle processes such as harassment and thermal evaporation may be at play in A2626. The presence of more \hi asymmetric galaxies in the substructures of A2626 indicates that tidal interactions between neighbouring galaxies likely pre-process the substructure galaxies before they fall into the cluster core. That the SF deficiency is not related to the \hi deficiency or \hi morphology of galaxies in all three environments suggests that the physical mechanisms causing disturbed \hi morphologies take some time to affect the star formation activity. Although it is important to investigate the \hi deficiencies and morphologies of galaxies, the interpretation is difficult without ancillary data that help us to better understand the effect of nurture on galaxy evolution.


\section{Summary}
\label{sec:summary}

We have presented \hi deficiencies, \hi morphologies, and star formation deficiencies of the \hi -detected galaxies in and around the A2626 cluster observed with \mk. There are three main environments in the volume we surveyed: non-substructure or isolated galaxies in A2626 (cluster environment), substructure galaxies in A2626 (groups influenced by the cluster environment), and the Swarm galaxies (group environment). We are interested in understanding whether the \hi deficiency, \hi morphology, and the star formation deficiency of the \hi -detected galaxies and the environment in which the galaxies reside are correlated. 

To characterise the \hi morphology of the outer \hi distribution of individual galaxies, we used three approaches. First, we used three visual classifications based on the outermost reliable \hi contour (settled, one-sided asymmetric, and unsettled galaxies). Second, we measured the offset of the \hi distribution with respect to the optical centre of the galaxy. Third, we calculated the modified asymmetry parameter \amod (as introduced by \citealt{Lelli2014}), which depends on the choice of the galaxy centre, the \hi column density above which it is measured, and how well the galaxy is spatially resolved. We chose an \hi column density threshold of 25 $\times 10^{19}$ cm$^{-2}$ to calculate the \amod value for galaxies with at least three beams in size and with a peak signal-to-noise ratio $\geq$ 5 in their \hi map. 

First, we explored the relation between our three different classification types of the \hi morphologies. We plotted \amod versus \hi offset from the optical centre and found a very strong correlation between them, as expected. We found that high \amod (\amod $\geq$ 0.4) low \hi offset galaxies are strongly \hi asymmetric galaxies. In the case of galaxies with both high \amod and high \hi offsets, the \amod value is driven by the large offset of the \hi disc compared to the stellar disc. Galaxies with a low \amod value and a small \hi offset are mostly \hi symmetric galaxies with settled \hi discs. There is a strong correspondence between \amod and our visual classifications, indicating that all three characterisations of \hi morphology are useful tracers of \hi asymmetries. 

In Fig. \ref{fig:HIdef_plot2} we investigated whether an environmental dependence exists on the \hi asymmetry and \hi offset. We find that substructures contain a higher fraction of asymmetric galaxies than the population of non-substructure galaxies in A2626. This suggests that tidal interactions may be more efficient inside substructures than outside substructures. This might be expected because the relative velocity differences of the neighbouring galaxies in the substructures tend to be lower than in the cluster galaxies outside the substructures.

There is a strong correlation between \hi deficiency and projected distance from the cluster centre. Outside R$_{200}$, non-substructure and substructure galaxies in A2626, and galaxies in the Swarm have a similar range of \hi deficiencies. This signifies that the mechanisms that cause \hi deficiencies might pre-process the galaxies prior to their infall into the cluster. Focusing on the most \hi deficient and most \hi rich galaxies in A2626, we observe that the \hi -deficient galaxies have yellowish optical colours and have small and truncated \hi discs, while the \hi -rich galaxies have bluish optical colours and extended \hi discs. This clearly shows that the \hi -deficient galaxies are in a later stage of their evolution, while \hi -rich galaxies are actively star forming with a sufficient amount of \hi gas.

Galaxies with an offset or asymmetric \hi disc are found in all three environments that we considered, implying that pre-processing of the \hi discs plays an important role. However, although the substructure galaxies in A2626 tend to be more \hi asymmetric than the non-substructure galaxies, the asymmetric galaxies are not necessarily \hi deficient. This suggests that although the substructure environment instigates more interactions among galaxies, these interactions do not completely deplete the \hi gas from these galaxies. 

The galaxy populations in all three different environments (substructure, non-substructure, and the Swarm) are mostly below the SFMS from \cite{cluver2020}. This means that the gas-removal mechanisms in the cluster environment as well as pre-processing in the substructures and the Swarm cause slightly lower SFRs than the usual SFR for normal galaxies. There is no clear relation between \hi deficiency, \hi asymmetry, or \hi offset with star formation deficiency for the galaxies in all the three environments. This signifies that the environmental mechanisms causing \hi deficiencies or asymmetries probably do not immediately impact the star formation activity, and vice versa.

We conclude that \hi deficiency and \hi morphology are good tracers of different environmental mechanisms acting on galaxies, and that their \hi characteristics and star formation deficiencies are affected by pre-processing before the galaxies enter the cluster environment. 

Encouraged by the high number of direct \hi detections through the unprecedented sensitivity of the \mk telescope, we will expand the \hi study of galaxies in and around A2626 by four additional \mk pointings in 32k-mode covering the ENE-SSW sector of A2626. The primary motivation of these observations is to enable a detailed investigation of so called pre-processing of \hi discs of galaxies before they enter the cluster environment. The higher spectral resolution of these observations will supplement the analysis of \hi morphologies presented in this paper, and will additionally allow for detailed investigations of the kinematical asymmetries of the detected galaxies.


\begin{acknowledgements}
We thank the anonymous referee for critical and constructive comments which helped improve this paper.
TD thanks Julia Healy for providing measurement of optical diameters of the galaxies. This paper makes use of the MeerKAT data (Project ID: SCI-20190418-JH-01). The MeerKAT telescope is operated by the South African Radio Astronomy Observatory, which is a facility of the National Research Foundation, an agency of the Department of Science and Innovation. MV acknowledges the Netherlands Foundation for Scientific Research support through VICI grant 639.043.511 and the Leids Kerkhoven-Bosscha Fonds (LKBF) for travel support. JMvdH acknowledges support from the European Research Council under the European Union’s Seventh Framework Programme (FP/2007-2013)/ERC Grant Agreement nr. 291531 (HiStoryNU)     
\end{acknowledgements}

%
%

\bibliographystyle{aa}
\bibliography{references}

\begin{thebibliography}{75}
\expandafter\ifx\csname natexlab\endcsname\relax\def\natexlab#1{#1}\fi

\bibitem[{Abramson {et~al.}(2011)Abramson, Kenney, Crowl, Chung, Van~Gorkom,
  Vollmer, \& Schiminovich}]{Abramson2011}
Abramson, A., Kenney, J.~D., Crowl, H.~H., {et~al.} 2011, Astronomical Journal,
  141

\bibitem[{Aguado {et~al.}(2019)Aguado, Ahumada, Almeida, Anderson, Andrews,
  Anguiano, Ort{\'{i}}z, Arag{\'{o}}n-Salamanca, Argudo-Fern{\'{a}}ndez,
  Aubert, Avila-Reese, Badenes, Barboza~Rembold, Barger, Barrera-Ballesteros,
  Bates, Bautista, Beaton, Beers, Belfiore, Bernardi, Bershady, Beutler, Bird,
  Bizyaev, Blanc, Blanton, Blomqvist, Bolton, Boquien, Borissova, Bovy,
  Nielsen~Brandt, Brinkmann, Brownstein, Bundy, Burgasser, Byler, Cano~Diaz,
  Cappellari, Carrera, Cervantes~Sodi, Chen, Cherinka, Doohyun~Choi, Chung,
  Coffey, Comerford, Comparat, Covey, da~Silva~Ilha, da~Costa, Sophia~Dai,
  Damke, Darling, Davies, Dawson, de~Sainte~Agathe, Deconto~Machado, Del~Moro,
  De~Lee, Diamond-Stanic, Dom{\'{i}}nguez~S{\'{a}}nchez, Donor, Drory,
  du~Mas~des Bourboux, Duckworth, Dwelly, Ebelke, Emsellem, Escoffier,
  Fern{\'{a}}ndez-Trincado, Feuillet, Fischer, Fleming, Fraser-McKelvie,
  Freischlad, Frinchaboy, Fu, Galbany, Garcia-Dias,
  Garc{\'{i}}a-Hern{\'{a}}ndez, Alberto Garma~Oehmichen, Antonio Geimba~Maia,
  Gil-Mar{\'{i}}n, Grabowski, Gu, Guo, Ha, Harrington, Hasselquist, Hayes,
  Hearty, Hernandez~Toledo, Hicks, Hogg, Holley-Bockelmann, Holtzman, Hsieh,
  Hunt, Seong~Hwang, Ibarra-Medel, Eduardo Jimenez~Angel, Johnson, Jones,
  J{\"{o}}nsson, Kinemuchi, Kollmeier, Krawczyk, Kreckel, Kruk, Lacerna, Lan,
  Lane, Law, Lee, Li, Lian, Lin, Lin, Lintott, Long, Longa-Pe{\~{n}}a,
  Ted~Mackereth, de~la Macorra, Majewski, Malanushenko, Manchado, Maraston,
  Mariappan, Marinelli, Marques-Chaves, Masseron, Masters, McDermid,
  Medina~Pe{\~{n}}a, Meneses-Goytia, Merloni, Merrifield, Meszaros, Minniti,
  Minsley, Muna, Myers, Nair, Correa~do Nascimento, Newman, Nitschelm,
  Olmstead, Oravetz, Oravetz, Ortega~Minakata, Pace, Padilla, Palicio, Pan,
  Pan, Parikh, Parker, Peirani, Penny, Percival, Perez-Fournon, Peterken,
  Pinsonneault, Prakash, Raddick, Raichoor, Riffel, Riffel, Rix, Robin,
  Roman-Lopes, Rose, Ross, Rossi, Rowlands, Rubin, S{\'{a}}nchez,
  S{\'{a}}nchez-Gallego, Sayres, Schaefer, Schiavon, Schimoia, Schlafly,
  Schlegel, Schneider, Schultheis, Seo, Shamsi, Shao, Shen, Shetty, Simonian,
  Smethurst, Sobeck, Souter, Spindler, Stark, Stassun, Steinmetz,
  Storchi-Bergmann, Stringfellow, Su{\'{a}}rez, Sun, Taghizadeh-Popp, Talbot,
  Tayar, Thakar, Thomas, Tissera, Tojeiro, Troup, Unda-Sanzana, Valenzuela,
  Vargas-Maga{\~{n}}a, Antonio V{\'{a}}zquez-Mata, Wake, Alan~Weaver, Weijmans,
  Westfall, Wild, Wilson, Woods, Yan, Yang, Zamora, Zasowski, Zhang, Zheng,
  Zheng, Zhu, Zinn, \& Zou}]{Aguado2018}
Aguado, D.~S., Ahumada, R., Almeida, A., {et~al.} 2019, The Astrophysical
  Journal Supplement Series, 240, 23

\bibitem[{Alpaslan {et~al.}(2016)Alpaslan, Grootes, Marcum, Popescu, Tuffs,
  Bland-Hawthorn, Brough, Brown, Davies, Driver, Holwerda, Kelvin,
  Lara-L{\'{o}}pez, L{\'{o}}pez-S{\'{a}}nchez, Loveday, Moffett, Taylor, Owers,
  \& Robotham}]{Alpaslan2016}
Alpaslan, M., Grootes, M., Marcum, P.~M., {et~al.} 2016, Monthly Notices of the
  Royal Astronomical Society, 457, 2287

\bibitem[{Baldry {et~al.}(2004)Baldry, Glazebrook, Brinkmann, Ivezi{\'{c}},
  Lupton, Nichol, \& Szalay}]{Baldry2004}
Baldry, I.~K., Glazebrook, K., Brinkmann, J., {et~al.} 2004, The Astrophysical
  Journal, 600, 681

\bibitem[{Balogh {et~al.}(2009)Balogh, McGee, Wilman, Bower, Hau, Morris,
  Mulchaey, Oemler, Parker, \& Gwyn}]{Balogh2009}
Balogh, M.~L., McGee, S.~L., Wilman, D., {et~al.} 2009, Monthly Notices of the
  Royal Astronomical Society, 398, 754

\bibitem[{{Balogh} {et~al.}(2000){Balogh}, {Navarro}, \& {Morris}}]{Balogh2000}
{Balogh}, M.~L., {Navarro}, J.~F., \& {Morris}, S.~L. 2000, \apj, 540, 113

\bibitem[{Batuski \& Burns(1985)}]{Batuski1985}
Batuski, D.~J. \& Burns, J.~O. 1985, Astrophysical Journal, 299, 5

\bibitem[{{Bigiel} {et~al.}(2008){Bigiel}, {Leroy}, {Walter}, {Brinks}, {de
  Blok}, {Madore}, \& {Thornley}}]{bigiel2008}
{Bigiel}, F., {Leroy}, A., {Walter}, F., {et~al.} 2008, aj, 136, 2846

\bibitem[{{Bilimogga} {et~al.}(2022){Bilimogga}, {Oman}, {Verheijen}, \& {van
  der Hulst}}]{bilimogga2022}
{Bilimogga}, P.~V., {Oman}, K.~A., {Verheijen}, M. A.~W., \& {van der Hulst},
  T. 2022, \mnras, 513, 5310

\bibitem[{{Bosma} {et~al.}(1977){Bosma}, {van der Hulst}, \&
  {Sullivan}}]{bosma1977}
{Bosma}, A., {van der Hulst}, J.~M., \& {Sullivan}, W.~T., I. 1977, aap, 57,
  373

\bibitem[{{Brinchmann} {et~al.}(2004){Brinchmann}, {Charlot}, {White},
  {Tremonti}, {Kauffmann}, {Heckman}, \& {Brinkmann}}]{brinchmann2004}
{Brinchmann}, J., {Charlot}, S., {White}, S.~D.~M., {et~al.} 2004, \mnras, 351,
  1151

\bibitem[{{Byrd} \& {Valtonen}(1990)}]{Byrd1990}
{Byrd}, G. \& {Valtonen}, M. 1990, \apj, 350, 89

\bibitem[{{Chamaraux} {et~al.}(1986){Chamaraux}, {Balkowski}, \&
  {Fontanelli}}]{chamaraux1986}
{Chamaraux}, P., {Balkowski}, C., \& {Fontanelli}, P. 1986, aap, 165, 15

\bibitem[{{Chung} {et~al.}(2009){Chung}, {van Gorkom}, {Kenney}, {Crowl}, \&
  {Vollmer}}]{chung2009}
{Chung}, A., {van Gorkom}, J.~H., {Kenney}, J. D.~P., {Crowl}, H., \&
  {Vollmer}, B. 2009, \aj, 138, 1741

\bibitem[{{Cluver} {et~al.}(2014){Cluver}, {Jarrett}, {Hopkins}, {Driver},
  {Liske}, {Gunawardhana}, {Taylor}, {Robotham}, {Alpaslan}, {Baldry}, {Brown},
  {Peacock}, {Popescu}, {Tuffs}, {Bauer}, {Bland-Hawthorn}, {Colless},
  {Holwerda}, {Lara-L{\'o}pez}, {Leschinski}, {L{\'o}pez-S{\'a}nchez},
  {Norberg}, {Owers}, {Wang}, \& {Wilkins}}]{cluver2014}
{Cluver}, M.~E., {Jarrett}, T.~H., {Hopkins}, A.~M., {et~al.} 2014, \apj, 782,
  90

\bibitem[{{Cluver} {et~al.}(2020){Cluver}, {Jarrett}, {Taylor}, {Hopkins},
  {Brough}, {Casura}, {Holwerda}, {Liske}, {Pimbblet}, \&
  {Wright}}]{cluver2020}
{Cluver}, M.~E., {Jarrett}, T.~H., {Taylor}, E.~N., {et~al.} 2020, \apj, 898,
  20

\bibitem[{{Conselice} {et~al.}(2003){Conselice}, {Bershady}, {Dickinson}, \&
  {Papovich}}]{conselice2003}
{Conselice}, C.~J., {Bershady}, M.~A., {Dickinson}, M., \& {Papovich}, C. 2003,
  \aj, 126, 1183

\bibitem[{{Cortese} {et~al.}(2021){Cortese}, {Catinella}, \&
  {Smith}}]{cortese2021}
{Cortese}, L., {Catinella}, B., \& {Smith}, R. 2021, pasa, 38, e035

\bibitem[{{Cowie} \& {Songaila}(1977)}]{Cowie1977}
{Cowie}, L.~L. \& {Songaila}, A. 1977, \nat, 266, 501

\bibitem[{{Crain} {et~al.}(2015){Crain}, {Schaye}, {Bower}, {Furlong},
  {Schaller}, {Theuns}, {Dalla Vecchia}, {Frenk}, {McCarthy}, {Helly},
  {Jenkins}, {Rosas-Guevara}, {White}, \& {Trayford}}]{crain2015}
{Crain}, R.~A., {Schaye}, J., {Bower}, R.~G., {et~al.} 2015, \mnras, 450, 1937

\bibitem[{Davies \& Lewis(1973)}]{Davies1973}
Davies, R.~D. \& Lewis, B.~M. 1973, Monthly Notices of the Royal Astronomical
  Society, 165, 231

\bibitem[{{Deb} {et~al.}(2020){Deb}, {Verheijen}, {Gullieuszik}, {Poggianti},
  {van Gorkom}, {Ramatsoku}, {Serra}, {Moretti}, {Vulcani}, {Bettoni},
  {Jaff{\'e}}, {Tonnesen}, \& {Fritz}}]{Deb2020}
{Deb}, T., {Verheijen}, M. A.~W., {Gullieuszik}, M., {et~al.} 2020, \mnras,
  494, 5029

\bibitem[{{Deb} {et~al.}(2022){Deb}, {Verheijen}, {Poggianti}, {Moretti}, {van
  der Hulst}, {Vulcani}, {Ramatsoku}, {Serra}, {Healy}, {Gullieuszik},
  {Bacchini}, {Ignesti}, {M{\"u}ller}, {Zabel}, {Luber}, {Jaff{\'e}}, \&
  {Gitti}}]{Deb2022}
{Deb}, T., {Verheijen}, M. A.~W., {Poggianti}, B.~M., {et~al.} 2022, arXiv
  e-prints, arXiv:2208.12950

\bibitem[{{D\'{e}nes} {et~al.}(2014){D\'{e}nes}, {Kilborn}, \&
  {Koribalski}}]{denes2014}
{D\'{e}nes}, H., {Kilborn}, V.~A., \& {Koribalski}, B.~S. 2014, mnras, 444, 667

\bibitem[{{Dey} {et~al.}(2019){Dey}, {Schlegel}, {Lang}, {Blum}, {Burleigh},
  {Fan}, {Findlay}, {Finkbeiner}, {Herrera}, {Juneau}, {Landriau}, {Levi},
  {McGreer}, {Meisner}, {Myers}, {Moustakas}, {Nugent}, {Patej}, {Schlafly},
  {Walker}, {Valdes}, {Weaver}, {Y{\`e}che}, {Zou}, {Zhou}, {Abareshi},
  {Abbott}, {Abolfathi}, {Aguilera}, {Alam}, {Allen}, {Alvarez}, {Annis},
  {Ansarinejad}, {Aubert}, {Beechert}, {Bell}, {BenZvi}, {Beutler}, {Bielby},
  {Bolton}, {Brice{\~n}o}, {Buckley-Geer}, {Butler}, {Calamida}, {Carlberg},
  {Carter}, {Casas}, {Castander}, {Choi}, {Comparat}, {Cukanovaite}, {Delubac},
  {DeVries}, {Dey}, {Dhungana}, {Dickinson}, {Ding}, {Donaldson}, {Duan},
  {Duckworth}, {Eftekharzadeh}, {Eisenstein}, {Etourneau}, {Fagrelius},
  {Farihi}, {Fitzpatrick}, {Font-Ribera}, {Fulmer}, {G{\"a}nsicke},
  {Gaztanaga}, {George}, {Gerdes}, {Gontcho}, {Gorgoni}, {Green}, {Guy},
  {Harmer}, {Hernandez}, {Honscheid}, {Huang}, {James}, {Jannuzi}, {Jiang},
  {Joyce}, {Karcher}, {Karkar}, {Kehoe}, {Kneib}, {Kueter-Young}, {Lan},
  {Lauer}, {Le Guillou}, {Le Van Suu}, {Lee}, {Lesser}, {Perreault Levasseur},
  {Li}, {Mann}, {Marshall}, {Mart{\'\i}nez-V{\'a}zquez}, {Martini}, {du Mas des
  Bourboux}, {McManus}, {Meier}, {M{\'e}nard}, {Metcalfe},
  {Mu{\~n}oz-Guti{\'e}rrez}, {Najita}, {Napier}, {Narayan}, {Newman}, {Nie},
  {Nord}, {Norman}, {Olsen}, {Paat}, {Palanque-Delabrouille}, {Peng},
  {Poppett}, {Poremba}, {Prakash}, {Rabinowitz}, {Raichoor}, {Rezaie},
  {Robertson}, {Roe}, {Ross}, {Ross}, {Rudnick}, {Safonova}, {Saha},
  {S{\'a}nchez}, {Savary}, {Schweiker}, {Scott}, {Seo}, {Shan}, {Silva},
  {Slepian}, {Soto}, {Sprayberry}, {Staten}, {Stillman}, {Stupak}, {Summers},
  {Sien Tie}, {Tirado}, {Vargas-Maga{\~n}a}, {Vivas}, {Wechsler}, {Williams},
  {Yang}, {Yang}, {Yapici}, {Zaritsky}, {Zenteno}, {Zhang}, {Zhang}, {Zhou}, \&
  {Zhou}}]{dey2019}
{Dey}, A., {Schlegel}, D.~J., {Lang}, D., {et~al.} 2019, \aj, 157, 168

\bibitem[{Dressler(1980)}]{Dressler1980}
Dressler, A. 1980, The Astrophysical Journal, 236, 351

\bibitem[{{Elbaz} {et~al.}(2007){Elbaz}, {Daddi}, {Le Borgne}, {Dickinson},
  {Alexander}, {Chary}, {Starck}, {Brandt}, {Kitzbichler}, {MacDonald},
  {Nonino}, {Popesso}, {Stern}, \& {Vanzella}}]{elbaz2007}
{Elbaz}, D., {Daddi}, E., {Le Borgne}, D., {et~al.} 2007, \aap, 468, 33

\bibitem[{{Giese} {et~al.}(2016){Giese}, {van der Hulst}, {Serra}, \&
  {Oosterloo}}]{giese2016}
{Giese}, N., {van der Hulst}, T., {Serra}, P., \& {Oosterloo}, T. 2016, \mnras,
  461, 1656

\bibitem[{{Gunn} \& {Gott}(1972)}]{Gunn1972}
{Gunn}, J.~E. \& {Gott}, III, J.~R. 1972, \apj, 176, 1

\bibitem[{{Han} {et~al.}(2018){Han}, {Smith}, {Choi}, {Cortese}, {Catinella},
  {Contini}, \& {Yi}}]{han2018}
{Han}, S., {Smith}, R., {Choi}, H., {et~al.} 2018, apj, 866, 78

\bibitem[{Haynes {et~al.}(1984)Haynes, Giovanelli, \& Chincarini}]{Haynes1984}
Haynes, M.~P., Giovanelli, R., \& Chincarini, G.~L. 1984, Annual Review of
  Astronomy and Astrophysics, 22, 445

\bibitem[{{Healy} {et~al.}(2021{\natexlab{a}}){Healy}, {Deb}, {Verheijen},
  {Blyth}, {Serra}, {Ramatsoku}, \& {Vulcani}}]{HD2021}
{Healy}, J., {Deb}, T., {Verheijen}, M.~A.~W., {et~al.} 2021{\natexlab{a}},
  arXiv e-prints, arXiv:2106.13018

\bibitem[{{Healy} {et~al.}(2021{\natexlab{b}}){Healy}, {Willner}, {Verheijen},
  \& {Blyth}}]{HealySS2021}
{Healy}, J., {Willner}, S.~P., {Verheijen}, M.~A.~W., \& {Blyth}, S.~L.
  2021{\natexlab{b}}, arXiv e-prints, arXiv:2106.08806

\bibitem[{{Holwerda} {et~al.}(2013){Holwerda}, {Pirzkal}, {de Blok}, \&
  {Blyth}}]{holwerda2013}
{Holwerda}, B.~W., {Pirzkal}, N., {de Blok}, W.~J.~G., \& {Blyth}, S.~L. 2013,
  mnras, 435, 1020

\bibitem[{{Holwerda} {et~al.}(2011){Holwerda}, {Pirzkal}, {de Blok},
  {Bouchard}, {Blyth}, {van der Heyden}, \& {Elson}}]{holwerda2011}
{Holwerda}, B.~W., {Pirzkal}, N., {de Blok}, W.~J.~G., {et~al.} 2011, \mnras,
  416, 2415

\bibitem[{Jaff{\'{e}} {et~al.}(2015)Jaff{\'{e}}, Smith, Candlish, Poggianti,
  Sheen, \& Verheijen}]{Jaffe2015}
Jaff{\'{e}}, Y.~L., Smith, R., Candlish, G.~N., {et~al.} 2015, Monthly Notices
  of the Royal Astronomical Society, 448, 1715

\bibitem[{{Jaff{\'e}} {et~al.}(2016){Jaff{\'e}}, {Verheijen}, {Haines}, {Yoon},
  {Cybulski}, {Montero-Casta{\~n}o}, {Smith}, {Chung}, {Deshev},
  {Fern{\'a}ndez}, {van Gorkom}, {Poggianti}, {Yun}, {Finoguenov}, {Smith}, \&
  {Okabe}}]{Jaffe2016}
{Jaff{\'e}}, Y.~L., {Verheijen}, M.~A.~W., {Haines}, C.~P., {et~al.} 2016,
  \mnras, 461, 1202

\bibitem[{{Kennicutt}(1989)}]{kennicutt1989}
{Kennicutt}, Robert~C., J. 1989, apj, 344, 685

\bibitem[{{Kennicutt}(1998)}]{Kennicutt1998}
{Kennicutt}, Jr., R.~C. 1998, araa, 36, 189

\bibitem[{Kleiner {et~al.}(2017)Kleiner, Pimbblet, Jones, Koribalsk, \&
  Serra}]{Kleiner2017}
Kleiner, D., Pimbblet, K.~A., Jones, D.~H., Koribalsk, B.~S., \& Serra, P. 2017

\bibitem[{Kraljic {et~al.}(2018)Kraljic, Arnouts, Pichon, Laigle, de~la Torre,
  Vibert, Cadiou, Dubois, Treyer, Schimd, Codis, de~Lapparent, Devriendt,
  Hwang, Le~Borgne, Malavasi, Milliard, Musso, Pogosyan, Alpaslan,
  Bland-Hawthorn, \& Wright}]{Kraljic2018}
Kraljic, K., Arnouts, S., Pichon, C., {et~al.} 2018, Monthly Notices of the
  Royal Astronomical Society, 474, 547

\bibitem[{Laigle {et~al.}(2018)Laigle, Pichon, Arnouts, McCracken, Dubois,
  Devriendt, Slyz, Le~Borgne, Benoit-L{\'{e}}vy, Hwang, Ilbert, Kraljic,
  Malavasi, Park, \& Vibert}]{Laigle2018}
Laigle, C., Pichon, C., Arnouts, S., {et~al.} 2018, Monthly Notices of the
  Royal Astronomical Society, 474, 5437

\bibitem[{{Larson} {et~al.}(1980){Larson}, {Tinsley}, \&
  {Caldwell}}]{Larson1980}
{Larson}, R.~B., {Tinsley}, B.~M., \& {Caldwell}, C.~N. 1980, \apj, 237, 692

\bibitem[{{Lelli} {et~al.}(2014){Lelli}, {Verheijen}, \&
  {Fraternali}}]{Lelli2014}
{Lelli}, F., {Verheijen}, M., \& {Fraternali}, F. 2014, mnras, 445, 1694

\bibitem[{{Loni} {et~al.}(2021){Loni}, {Serra}, {Kleiner}, {Cortese},
  {Catinella}, {Koribalski}, {Jarrett}, {Molnar}, {Davis}, {Iodice},
  {Lee-Waddell}, {Loi}, {Maccagni}, {Peletier}, {Popping}, {Ramatsoku},
  {Smith}, \& {Zabel}}]{loni2021}
{Loni}, A., {Serra}, P., {Kleiner}, D., {et~al.} 2021, \aap, 648, A31

\bibitem[{{Lotz} {et~al.}(2004){Lotz}, {Primack}, \& {Madau}}]{lotz2004}
{Lotz}, J.~M., {Primack}, J., \& {Madau}, P. 2004, \aj, 128, 163

\bibitem[{{McGee} {et~al.}(2009){McGee}, {Balogh}, {Bower}, {Font}, \&
  {McCarthy}}]{mcgee2009}
{McGee}, S.~L., {Balogh}, M.~L., {Bower}, R.~G., {Font}, A.~S., \& {McCarthy},
  I.~G. 2009, mnras, 400, 937

\bibitem[{{Merritt}(1983)}]{Merritt1983}
{Merritt}, D. 1983, \apj, 264, 24

\bibitem[{{Molnar} {et~al.}(2021){Molnar}, {Serra}, {van der Hulst}, {Jarrett},
  {Boselli}, {Cortese}, {Healy}, {de Blok}, {Cappellari}, {Hess}, {Jozsa},
  {McDermid}, {Oosterloo}, \& {Verheijen}}]{Molnar2021}
{Molnar}, D.~C., {Serra}, P., {van der Hulst}, T., {et~al.} 2021, arXiv
  e-prints, arXiv:2112.12244

\bibitem[{{Moore} {et~al.}(1996){Moore}, {Katz}, {Lake}, {Dressler}, \&
  {Oemler}}]{Moore1996}
{Moore}, B., {Katz}, N., {Lake}, G., {Dressler}, A., \& {Oemler}, A. 1996,
  \nat, 379, 613

\bibitem[{Nelson(1982)}]{Nelson1982}
Nelson, A.~H. 1982, Nature, 295, 263

\bibitem[{{Noeske} {et~al.}(2007){Noeske}, {Weiner}, {Faber}, {Papovich},
  {Koo}, {Somerville}, {Bundy}, {Conselice}, {Newman}, {Schiminovich}, {Le
  Floc'h}, {Coil}, {Rieke}, {Lotz}, {Primack}, {Barmby}, {Cooper}, {Davis},
  {Ellis}, {Fazio}, {Guhathakurta}, {Huang}, {Kassin}, {Martin}, {Phillips},
  {Rich}, {Small}, {Willmer}, \& {Wilson}}]{Noeske2007}
{Noeske}, K.~G., {Weiner}, B.~J., {Faber}, S.~M., {et~al.} 2007, \apjl, 660,
  L43

\bibitem[{{Poggianti} {et~al.}(2016){Poggianti}, {Fasano}, {Omizzolo},
  {Gullieuszik}, {Bettoni}, {Moretti}, {Paccagnella}, {Jaff{\'e}}, {Vulcani},
  {Fritz}, {Couch}, \& {D'Onofrio}}]{Poggianti2016}
{Poggianti}, B.~M., {Fasano}, G., {Omizzolo}, A., {et~al.} 2016, \aj, 151, 78

\bibitem[{{Poggianti} {et~al.}(2017){Poggianti}, {Moretti}, {Gullieuszik},
  {Fritz}, {Jaff{\'e}}, {Bettoni}, {Fasano}, {Bellhouse}, {Hau}, {Vulcani},
  {Biviano}, {Omizzolo}, {Paccagnella}, {D'Onofrio}, {Cava}, {Sheen}, {Couch},
  \& {Owers}}]{Poggianti2017}
{Poggianti}, B.~M., {Moretti}, A., {Gullieuszik}, M., {et~al.} 2017, \apj, 844,
  48

\bibitem[{{Ramatsoku} {et~al.}(2020){Ramatsoku}, {Serra}, {Poggianti},
  {Moretti}, {Gullieuszik}, {Bettoni}, {Deb}, {Franchetto}, {van Gorkom},
  {Jaff{\'e}}, {Tonnesen}, {Verheijen}, {Vulcani}, {Andati}, {de Blok},
  {J{\'o}zsa}, {Kamphuis}, {Kleiner}, {Maccagni}, {Makhathini}, {Moln{\'a}r},
  {Ramaila}, {Smirnov}, \& {Thorat}}]{ramatsoku2020}
{Ramatsoku}, M., {Serra}, P., {Poggianti}, B.~M., {et~al.} 2020, \aap, 640, A22

\bibitem[{{Ramatsoku} {et~al.}(2019){Ramatsoku}, {Serra}, {Poggianti},
  {Moretti}, {Gullieuszik}, {Bettoni}, {Deb}, {Fritz}, {van Gorkom}, \&
  {Jaffe}}]{ramatsoku2019}
{Ramatsoku}, M., {Serra}, P., {Poggianti}, B.~M., {et~al.} 2019, \mnras, 487,
  4580

\bibitem[{Sarron {et~al.}(2018)Sarron, Martinet, Durret, \& Adami}]{Sarron2018}
Sarron, F., Martinet, N., Durret, F., \& Adami, C. 2018, Astronomy and
  Astrophysics, 613, 67

\bibitem[{{Schaye} {et~al.}(2015){Schaye}, {Crain}, {Bower}, {Furlong},
  {Schaller}, {Theuns}, {Dalla Vecchia}, {Frenk}, {McCarthy}, {Helly},
  {Jenkins}, {Rosas-Guevara}, {White}, {Baes}, {Booth}, {Camps}, {Navarro},
  {Qu}, {Rahmati}, {Sawala}, {Thomas}, \& {Trayford}}]{schaye2015}
{Schaye}, J., {Crain}, R.~A., {Bower}, R.~G., {et~al.} 2015, \mnras, 446, 521

\bibitem[{Serra {et~al.}(2013)Serra, Koribalski, Duc, Oosterloo, McDermid,
  Michel-Dansac, Emsellem, Cuillandre, Alatalo, Blitz, Bois, Bournaud, Bureau,
  Cappellari, Crocker, Davies, Davis, Zeeuw, Khochfar, Krajnovi{\'{c}},
  Kuntschner, Lablanche, Morganti, Naab, Sarzi, Scott, Weijmans, \&
  Young}]{Serra2013}
Serra, P., Koribalski, B., Duc, P.~A., {et~al.} 2013, Monthly Notices of the
  Royal Astronomical Society, 428, 370

\bibitem[{Serra {et~al.}(2015)Serra, Westmeier, Giese, Jurek, Fl{\"{o}}er,
  Popping, Winkel, Van~der Hulst, Meyer, Koribalski, Staveley-Smith, \&
  Courtois}]{Serra2015}
Serra, P., Westmeier, T., Giese, N., {et~al.} 2015, Monthly Notices of the
  Royal Astronomical Society, 448, 1922

\bibitem[{{Solanes} {et~al.}(1996){Solanes}, {Giovanelli}, \&
  {Haynes}}]{solanes1996}
{Solanes}, J.~M., {Giovanelli}, R., \& {Haynes}, M.~P. 1996, apj, 461, 609

\bibitem[{Solanes {et~al.}(2001)Solanes, Manrique, Garc{\'{i}}a‐G{\'{o}}mez,
  Gonz{\'{a}}lez‐Casado, Giovanelli, \& Haynes}]{Solanes2001}
Solanes, J.~M., Manrique, A., Garc{\'{i}}a‐G{\'{o}}mez, C., {et~al.} 2001,
  The Astrophysical Journal, 548, 97

\bibitem[{{Speagle} {et~al.}(2014){Speagle}, {Steinhardt}, {Capak}, \&
  {Silverman}}]{speagle2014}
{Speagle}, J.~S., {Steinhardt}, C.~L., {Capak}, P.~L., \& {Silverman}, J.~D.
  2014, \apjs, 214, 15

\bibitem[{{Spitzer} \& {Baade}(1951)}]{Spitzer1951}
{Spitzer}, Jr., L. \& {Baade}, W. 1951, \apj, 113, 413

\bibitem[{{Springel}(2000)}]{Springel2000}
{Springel}, V. 2000, \mnras, 312, 859

\bibitem[{{Tinsley} \& {Larson}(1979)}]{Tinsley1979}
{Tinsley}, B.~M. \& {Larson}, R.~B. 1979, \mnras, 186, 503

\bibitem[{{Tomczak} {et~al.}(2016){Tomczak}, {Quadri}, {Tran}, {Labb{\'e}},
  {Straatman}, {Papovich}, {Glazebrook}, {Allen}, {Brammer}, {Cowley},
  {Dickinson}, {Elbaz}, {Inami}, {Kacprzak}, {Morrison}, {Nanayakkara},
  {Persson}, {Rees}, {Salmon}, {Schreiber}, {Spitler}, \&
  {Whitaker}}]{tomczak2016}
{Tomczak}, A.~R., {Quadri}, R.~F., {Tran}, K.-V.~H., {et~al.} 2016, \apj, 817,
  118

\bibitem[{Tonnesen {et~al.}(2007)Tonnesen, Bryan, \& van Gorkom}]{Tonnesen2007}
Tonnesen, S., Bryan, G.~L., \& van Gorkom, J.~H. 2007, The Astrophysical
  Journal, 671, 1434

\bibitem[{Toomre \& Toomre(1972)}]{Toomre1972}
Toomre, A. \& Toomre, J. 1972, The Astrophysical Journal, 178, 623

\bibitem[{{Valluri}(1993)}]{Valluri1993}
{Valluri}, M. 1993, \apj, 408, 57

\bibitem[{Vulcani {et~al.}(2019)Vulcani, Poggianti, Moretti, Gullieuszik,
  Fritz, Franchetto, Fasano, Bettoni, \& Jaff{\'{e}}}]{Vulcani2019}
Vulcani, B., Poggianti, B.~M., Moretti, A., {et~al.} 2019, Monthly Notices of
  the Royal Astronomical Society, 487, 2278

\bibitem[{Williams \& Rood(1987)}]{Williams1987}
Williams, B.~A. \& Rood, H.~J. 1987, The Astrophysical Journal Supplement
  Series, 63, 265

\bibitem[{Yoon {et~al.}(2017)Yoon, Chung, Smith, \& Jaff{\'{e}}}]{Yoon2017}
Yoon, H., Chung, A., Smith, R., \& Jaff{\'{e}}, Y.~L. 2017, The Astrophysical
  Journal, 838, 81

\bibitem[{York {et~al.}(2000)York, Adelman, John E.~Anderson, Anderson, Annis,
  Bahcall, Bakken, Barkhouser, Bastian, Berman, Boroski, Bracker, Briegel,
  Briggs, Brinkmann, Brunner, Burles, Carey, Carr, Castander, Chen, Colestock,
  Connolly, Crocker, Csabai, Czarapata, Davis, Doi, Dombeck, Eisenstein,
  Ellman, Elms, Evans, Fan, Federwitz, Fiscelli, Friedman, Frieman, Fukugita,
  Gillespie, Gunn, Gurbani, Haas, Haldeman, Harris, Hayes, Heckman, Hennessy,
  Hindsley, Holm, Holmgren, Huang, Hull, Husby, Ichikawa, Ichikawa,
  Ivezi{\'{c}}, Kent, Kim, Kinney, Klaene, Kleinman, Kleinman, Knapp, Korienek,
  Kron, Kunszt, Lamb, Lee, Leger, Limmongkol, Lindenmeyer, Long, Loomis,
  Loveday, Lucinio, Lupton, MacKinnon, Mannery, Mantsch, Margon, McGehee,
  McKay, Meiksin, Merelli, Monet, Munn, Narayanan, Nash, Neilsen, Neswold,
  Newberg, Nichol, Nicinski, Nonino, Okada, Okamura, Ostriker, Owen, Pauls,
  Peoples, Peterson, Petravick, Pier, Pope, Pordes, Prosapio, Rechenmacher,
  Quinn, Richards, Richmond, Rivetta, Rockosi, Ruthmansdorfer, Sandford,
  Schlegel, Schneider, Sekiguchi, Sergey, Shimasaku, Siegmund, Smee, Smith,
  Snedden, Stone, Stoughton, Strauss, Stubbs, SubbaRao, Szalay, Szapudi,
  Szokoly, Thakar, Tremonti, Tucker, Uomoto, Berk, Vogeley, Waddell, Wang,
  Watanabe, Weinberg, Yanny, \& Yasuda}]{York2000}
York, D.~G., Adelman, J., John E.~Anderson, J., {et~al.} 2000, The Astronomical
  Journal, 120, 1579

\bibitem[{{Zaragoza-Cardiel} {et~al.}(2018){Zaragoza-Cardiel}, {Smith},
  {Rosado}, {Beckman}, {Bitsakis}, {Camps-Fari{\~n}a}, {Font}, \&
  {Cox}}]{Zaragoza-Cardie2018}
{Zaragoza-Cardiel}, J., {Smith}, B.~J., {Rosado}, M., {et~al.} 2018, apjs, 234,
  35

\end{thebibliography}

\begin{appendix}
\onecolumn
\section{Catalogue of \hi detections}

\centering
\setlength{\tabcolsep}{6pt} 
\renewcommand{\arraystretch}{1}

\centering
\setlength{\tabcolsep}{3pt} 
\renewcommand{\arraystretch}{1}

\begin{longtable}[t]{ccccccccccccccc}
\caption[]{List of all of the relevant \hi and stellar physical and morphological characteristics of the galaxies. The column entries of table \ref{tab:catalogue} are mentioned at the end of the table.} \\\hline
\multicolumn{1}{c}{HI ID} &
\multicolumn{1}{c}{Name} &
\multicolumn{1}{c}{SS} &
\multicolumn{1}{c}{z} &
\multicolumn{1}{c}{Log(M$\rm_{HI}$}) &
\multicolumn{1}{c}{$\pm$} &
\multicolumn{1}{c}{HIdef} &
\multicolumn{1}{c}{Amod} &
\multicolumn{1}{c}{\hi offset} &
\multicolumn{1}{c}{VClass} &
\multicolumn{1}{c}{Log(M$_{*}$}) &
\multicolumn{1}{c}{$\pm$} &
\multicolumn{1}{c}{SFR} &
\multicolumn{1}{c}{$\pm$} &
\multicolumn{1}{c}{g-r} \\
& & 
\multicolumn{2}{c}{} &
\multicolumn{2}{c}{} &
\multicolumn{2}{c}{} &
\multicolumn{1}{c}{(kpc)} &
\multicolumn{1}{c}{} &
\multicolumn{2}{c}{} &
\multicolumn{2}{c}{(M$_{\odot}$/yr)} &
\multicolumn{1}{c}{mag}\\
\multicolumn{1}{c}{(1)}&
\multicolumn{1}{c}{(2)}&
\multicolumn{1}{c}{(3)}&
\multicolumn{1}{c}{(4)}&
\multicolumn{1}{c}{(5)}&
\multicolumn{1}{c}{(6)}&
\multicolumn{1}{c}{(7)}&
\multicolumn{1}{c}{(8)}&
\multicolumn{1}{c}{(9)}&
\multicolumn{1}{c}{(10)}&
\multicolumn{1}{c}{(11)}&
\multicolumn{1}{c}{(12)}&
\multicolumn{1}{c}{(13)}&
\multicolumn{1}{c}{(14)}&
\multicolumn{1}{c}{(15)}\\

\hline
\hline
\endfirsthead

\noalign{\vspace{0.7mm}}
\multicolumn{15}{l}{Continued from previous page.} \\
\hline
\multicolumn{1}{c}{HI ID} &
\multicolumn{1}{c}{Name} &
\multicolumn{1}{c}{SS} &
\multicolumn{1}{c}{z} &
\multicolumn{1}{c}{Log(M$\rm_{HI}$)} &
\multicolumn{1}{c}{$\pm$} &
\multicolumn{1}{c}{HIdef} &
\multicolumn{1}{c}{Amod} &
\multicolumn{1}{c}{\hi offset} &
\multicolumn{1}{c}{VClass} &
\multicolumn{1}{c}{Log(M$_{*}$)} &
\multicolumn{1}{c}{$\pm$} &
\multicolumn{1}{c}{SFR} &
\multicolumn{1}{c}{$\pm$} &
\multicolumn{1}{c}{g-r} \\
& & 
\multicolumn{2}{c}{} &
\multicolumn{2}{c}{} &
\multicolumn{2}{c}{} &
\multicolumn{1}{c}{(kpc)} &
\multicolumn{1}{c}{} &
\multicolumn{2}{c}{} &
\multicolumn{2}{c}{(M$_{\odot}$/yr)} &
\multicolumn{1}{c}{mag}\\
\multicolumn{1}{c}{(1)}&
\multicolumn{1}{c}{(2)}&
\multicolumn{1}{c}{(3)}&
\multicolumn{1}{c}{(4)}&
\multicolumn{1}{c}{(5)}&
\multicolumn{1}{c}{(6)}&
\multicolumn{1}{c}{(7)}&
\multicolumn{1}{c}{(8)}&
\multicolumn{1}{c}{(9)}&
\multicolumn{1}{c}{(10)}&
\multicolumn{1}{c}{(11)}&
\multicolumn{1}{c}{(12)}&
\multicolumn{1}{c}{(13)}&
\multicolumn{1}{c}{(14)}&
\multicolumn{1}{c}{(15)}\\
\hline
\hline
\endhead
\endfoot
\endlastfoot
10 & J233409.36+211641.9 & 1 & 0.0524 & 9.24 & 0.09 & 0.12 & 0.60 & 2.79 & 2 & 8.83 & 1.28 & 0.23 & -.-- & 0.42\\
  12 & J233413.05+212327.5 & 1 & 0.0515 & 9.60 & 0.05 & 0.07 & 0.48 & 2.75 & 2 & 9.52 & 0.13 & 0.55 & 0.22 & 0.55\\
  13 & J233425.70+213122.9 & 1 & 0.0553 & 9.74 & 0.04 & -0.02 & 0.43 & 3.35 & 2 & 9.56 & 1.27 & 0.37 & -.-- & 0.46\\
  15 & J233438.15+211851.7 & 1 & 0.0538 & 9.93 & 0.03 & -0.07 & 0.45 & 2.06 & 2 & 10.36 & 0.15 & 3.33 & 1.15 & 0.73\\
  16 & J233438.80+211721.0 & 1 & 0.0529 & 9.07 & 0.08 & 0.64 & 0.87 & 5.85 & 2 & 10.22 & 0.11 & 2.0 & 0.70 & 0.73\\
  17 & J233440.31+203710.8 & 0 & 0.0573 & 9.95 & 0.04 & 0.01 & 0.64 & 6.18 & 2 & 10.46 & 0.10 & 3.08 & 1.07 & 0.70\\
  19 & J233453.14+213344.9 & 1 & 0.0554 & 9.38 & 0.06 & 0.39 & 0.49 & 0.85 & 2 & 9.78 & 0.13 & 2.03 & 0.71 & 0.77\\
  21 & J233455.82+212245.8 & 1 & 0.0525 & 9.50 & 0.04 & 0.21 & 0.54 & 3.38 & 2 & 9.63 & 1.27 & 0.36 & -.-- & 0.52\\
  23 & J233500.37+205908.6 & 0 & 0.0588 & 9.31 & 0.06 & 0.34 & 0.39 & 1.80 & 1 & 9.18 & 1.27 & 0.27 & -.-- & 0.40\\
  24 & J233510.54+212147.8 & 1 & 0.0527 & 9.35 & 0.05 & 0.42 & 0.37 & 0.93 & 2 & 9.99 & 0.26 & 0.74 & 0.28 & 0.63\\
  25 & J233512.34+214628.1 & 2 & 0.0649 & 9.58 & 0.08 & -.-- & 0.29 & 0.79 & 1 & 10.23 & 0.2 & 2.14 & 0.76 & 0.60\\
  29 & J233525.93+204419.5 & 2 & 0.0617 & 9.26 & 0.09 & 0.07 & 0.34 & 2.16 & 1 & 8.81 & 1.27 & 0.31 & -.-- & 0.37\\
  31 & J233526.81+211638.3 & 2 & 0.0650 & 9.87 & 0.03 & -0.23 & 0.35 & 0.31 & 2 & 10.12 & 0.14 & 1.34 & 0.48 & 0.83\\
  32 & J233527.64+204059.2 & 2 & 0.0662 & 9.54 & 0.09 & -0.36 & 0.87 & 37.91 & 1 & 8.85 & 1.29 & 0.33 & -.-- & 0.36\\
  33 & J233528.63+203803.2 & 2 & 0.0625 & 9.54 & 0.09 & -0.04 & 0.21 & 1.10 & 1 & 9.22 & 1.27 & 0.42 & -.-- & 0.29\\
  36 & J233532.73+211011.3 & 2 & 0.0659 & 10.21 & 0.01 & -0.51 & 0.35 & 1.31 & 1 & 9.91 & 0.31 & 1.01 & 0.39 & 0.46\\
  37 & J233533.49+210252.1 & 0 & 0.0567 & 9.63 & 0.04 & 0.13 & 0.46 & 2.08 & 2 & 10.40 & 0.24 & 2.19 & 0.77 & 0.71\\
  39 & J233535.77+204159.6 & 1 & 0.0613 & 9.99 & 0.04 & -0.10 & 0.50 & 2.56 & 3 & 10.72 & 0.10 & 3.88 & 1.35 & 0.78\\
  40 & J233535.93+203807.1 & 2 & 0.0626 & 10.02 & 0.03 & -0.51 & 0.29 & 2.50 & 1 & 10.12 & 0.26 & 0.37 & 0.17 & 0.48\\
  42 & J233536.98+210440.3 & 0 & 0.0545 & 9.23 & 0.04 & -0.04 & 0.33 & 1.93 & 2 & 9.44 & 1.27 & 0.35 & -.-- & 0.54\\
  43 & J233537.03+204639.6 & 0 & 0.0486 & 9.21 & 0.06 & 0.46 & 0.47 & 1.50 & 2 & 9.33 & 1.27 & 0.29 & -.-- & 0.53\\
  44 & J233537.44+211025.1 & 2 & 0.0654 & 9.12 & 0.08 & 0.17 & 0.86 & 1.75 & 2 & 8.99 & 1.29 & 0.39 & -.-- & 0.29\\
  45 & J233539.99+210844.6 & 0 & 0.0613 & 9.81 & 0.02 & -0.17 & 0.25 & 1.05 & 3 & 9.54 & 1.27 & 0.26 & -.-- & 0.49\\
  46 & J233540.43+205357.4 & 0 & 0.0578 & 8.90 & 0.11 & -.-- & 0.97 & 1.94 & 1 & -.-- & -.-- & -.-- & -.-- & 0.29\\
  52 & J233544.62+210228.7 & 2 & 0.0663 & 9.55 & 0.04 & 0.18 & 0.13 & 0.54 & 1 & 9.69 & 0.25 & 0.73 & 0.29 & 0.62\\
  53 & J233545.43+205045.5 & 2 & 0.0639 & 9.10 & 0.09 & 0.13 & 0.64 & 2.15 & 2 & -.-- & -.-- & -.-- & -.-- & 0.47\\
  54 & J233546.25+210031.0 & 1 & 0.0540 & 9.34 & 0.04 & -0.04 & 0.30 & 0.98 & 1 & 8.63 & 1.28 & 0.27 & -.-- & 0.32\\
  55 & J233546.32+203026.4 & 2 & 0.0655 & 10.16 & 0.04 & -0.48 & 0.57 & 0.75 & 3 & 10.34 & 0.14 & 2.07 & 0.73 & 0.59\\
  57 & J233547.60+210207.0 & 2 & 0.0636 & 9.28 & 0.05 & 0.39 & 0.72 & 3.94 & 3 & 10.38 & 0.13 & 2.38 & 0.84 & 0.72\\
  58 & J233547.85+204226.6 & 2 & 0.0619 & 10.02 & 0.03 & 0.24 & 0.29 & 3.37 & 1 & 11.23 & 0.10 & 5.04 & 1.76 & 0.87\\
  59 & J233549.79+211356.4 & 2 & 0.0661 & 8.96 & 0.08 & 0.43 & 1.00 & 3.91 & 2 & 9.47 & 1.14 & 0.4 & 0.2 & 0.44\\
  60 & J233550.08+202318.3 & 2 & 0.0639 & 10.09 & 0.05 & -.-- & 0.74 & 7.28 & 3 & 10.54 & 0.12 & 2.75 & 0.97 & 0.82\\
  62 & J233556.63+211333.0 & 2 & 0.0660 & 9.53 & 0.04 & 0.01 & 0.13 & 0.24 & 1 & 9.68 & 1.27 & 0.29 & -.-- & 0.44\\
  63 & J233557.05+211105.5 & 2 & 0.0649 & 9.94 & 0.03 & -0.06 & 0.33 & 1.10 & 1 & 10.42 & 0.13 & 3.80 & 1.33 & 1.17\\
  64 & J233557.38+205307.1 & 0 & 0.0597 & 9.42 & 0.05 & 0.31 & 0.55 & 2.61 & 2 & 9.89 & 0.35 & 0.41 & 0.19 & 0.60\\
  65 & J233557.68+211703.5 & 0 & 0.0501 & 8.99 & 0.06 & 0.71 & 0.22 & 0.92 & 1 & 9.44 & 0.38 & 0.26 & 0.14 & 0.53\\
  70 & J233602.82+210821.1 & 2 & 0.0644 & 9.35 & 0.05 & -.-- & 0.30 & 0.98 & 1 & 9.81 & 0.27 & 0.72 & 0.31 & 0.26\\
  71 & J233604.52+210613.3 & 2 & 0.0661 & 9.65 & 0.03 & 0.25 & 0.90 & 4.83 & 3 & 10.77 & 0.10 & 0.24 & 0.13 & 0.92\\
  73 & J233606.20+203113.8 & 2 & 0.0650 & 9.61 & 0.07 & -0.19 & 0.54 & 3.40 & 2 & 8.98 & 1.27 & 0.34 & -.-- & 0.38\\
  75 & J233607.00+210500.1 & 0 & 0.0576 & 9.44 & 0.04 & 0.00 & 0.27 & 0.56 & 1 & 8.89 & 1.27 & 0.30 & -.-- & 0.30\\
  77 & J233609.55+205428.4 & 0 & 0.0492 & 9.05 & 0.06 & -.-- & 0.83 & 2.78 & 1 & -.-- & -.-- & -.-- & -.-- & 0.32\\
  79 & J233611.37+205702.0 & 0 & 0.0536 & 9.22 & 0.06 & 0.10 & 0.56 & 4.16 & 2 & -.-- & -.-- & -.-- & -.-- & 0.28\\
  81 & J233614.66+214434.3 & 1 & 0.0560 & 9.71 & 0.04 & -0.11 & 0.45 & 2.75 & 2 & 9.70 & 1.27 & 0.50 & -.-- & 0.37\\
  84 & J233615.59+205047.1 & 0 & 0.0492 & 9.27 & 0.05 & 0.19 & 0.52 & 1.97 & 2 & 9.32 & 0.28 & 0.52 & 0.20 & 0.38\\
  86 & J233616.89+215412.1 & 1 & 0.0565 & 9.35 & 0.10 & -0.11 & 0.37 & 0.93 & 1 & 9.12 & 1.27 & 0.24 & -.-- & 0.23\\
  87 & J233618.06+203054.6 & 2 & 0.0649 & 10.15 & 0.03 & -0.25 & 0.49 & 2.45 & 3 & 10.37 & 0.12 & 1.90 & 0.68 & 0.94\\
  88 & J233618.57+210402.6 & 0 & 0.0533 & 9.77 & 0.02 & 0.20 & 0.56 & 3.40 & 2 & 9.89 & 0.14 & 2.31 & 0.81 & 0.47\\
  90 & J233619.79+204950.3 & 0 & 0.0514 & 8.84 & 0.12 & 0.60 & 1.00 & 4.29 & 2 & 9.53 & 1.27 & 0.36 & -.-- & 0.48\\
  95 & J233624.23+204355.5 & 2 & 0.0634 & 9.22 & 0.07 & 0.21 & 0.49 & 1.43 & 1 & 9.34 & 1.27 & 0.31 & -.-- & 0.38\\
  99 & J233626.71+215204.9 & 1 & 0.0570 & 10.26 & 0.03 & -0.38 & 0.69 & 6.67 & 3 & 9.90 & 0.14 & 1.10 & 0.40 & 0.06\\
  102 & J233631.03+205750.0 & 0 & 0.0585 & 9.21 & 0.07 & 0.49 & 0.70 & 6.88 & 2 & 10.03 & 0.14 & 2.02 & 0.71 & 0.99\\
  103 & J233631.53+205325.7 & 2 & 0.0624 & 9.10 & 0.07 & 0.59 & 1.00 & 5.30 & 2 & 10.05 & 0.14 & 1.66 & 0.59 & 0.61\\
  105 & J233632.47+204110.8 & 2 & 0.0646 & 9.98 & 0.04 & -.-- & 0.77 & 6.74 & 2 & 10.24 & 0.10 & 5.75 & 1.99 & 0.81\\
  108 & J233632.86+203424.6 & 2 & 0.0661 & 9.13 & 0.09 & 0.57 & 0.88 & 4.97 & 2 & 10.18 & 0.11 & 5.29 & 1.84 & 0.81\\
  109 & J233634.21+203441.1 & 2 & 0.0673 & 9.55 & 0.05 & 0.22 & 0.29 & 0.44 & 2 & 10.22 & 0.10 & 6.65 & 2.31 & 0.59\\
  110 & J233639.74+210606.8 & 0 & 0.0589 & 9.46 & 0.04 & 0.65 & 0.57 & 3.86 & 2 & 10.98 & 0.11 & 6.35 & 2.20 & 0.74\\
  111 & J233642.05+212810.6 & 0 & 0.0531 & 9.33 & 0.05 & 0.26 & 0.68 & 3.28 & 2 & 9.88 & 0.11 & 0.69 & 0.26 & 0.45\\
  113 & J233642.70+212212.0 & 0 & 0.0550 & 9.58 & 0.03 & 0.16 & 0.39 & 2.17 & 1 & 10.08 & 0.13 & 2.52 & 0.88 & 0.63\\
  115 & J233645.05+205108.7 & 0 & 0.0530 & 9.28 & 0.05 & -0.21 & 0.90 & 5.14 & 3 & 9.06 & 0.52 & 0.37 & 0.17 & 0.56\\
  117 & J233647.54+212506.6 & 0 & 0.0563 & 9.02 & 0.10 & 0.29 & 1.00 & 5.84 & 2 & -.-- & -.-- & -.-- & -.-- & 0.40\\
  118 & J233647.73+204136.1 & 0 & 0.0539 & 9.91 & 0.02 & 0.11 & 0.28 & 1.65 & 1 & 10.18 & 0.11 & 1.27 & 0.45 & 0.61\\
  119 & J233648.56+205031.3 & 0 & 0.0521 & 8.77 & 0.12 & 0.56 & 0.41 & 1.47 & 1 & 8.98 & 1.28 & 0.34 & -.-- & 0.25\\
  122 & J233650.53+203534.6 & 0 & 0.0571 & 9.88 & 0.04 & -0.10 & 0.68 & 6.92 & 3 & 9.69 & 1.27 & 0.23 & -.-- & 0.51\\
  124 & J233652.31+204719.4 & 0 & 0.0592 & 9.39 & 0.06 & 0.29 & 0.30 & 0.86 & 1 & 9.70 & 1.27 & 0.47 & -.-- & 0.56\\
  126 & J233652.59+203726.2 & 0 & 0.0506 & 9.43 & 0.06 & 0.31 & 0.54 & 3.68 & 2 & 9.11 & 1.27 & 0.39 & -.-- & 0.44\\
  127 & J233654.75+203635.9 & 0 & 0.0517 & 9.23 & 0.08 & 0.24 & 0.49 & 3.24 & 2 & 9.05 & 1.27 & 0.4 & -.-- & 0.28\\
  128 & J233655.10+212708.9 & 0 & 0.0528 & 9.46 & 0.05 & 0.67 & 0.52 & 2.75 & 2 & 10.99 & 0.21 & 4.43 & 1.54 & 0.79\\
  129 & J233655.43+204826.1 & 0 & 0.0579 & 9.94 & 0.02 & 0.09 & 0.30 & 1.05 & 1 & 10.02 & 0.16 & 1.35 & 0.48 & 0.52\\
  131 & J233657.81+204254.2 & 2 & 0.0633 & 9.57 & 0.05 & -0.05 & 0.51 & 2.57 & 3 & 9.56 & 1.27 & 0.48 & -.-- & 0.40\\
  132 & J233658.57+203442.4 & 0 & 0.0517 & 9.72 & 0.04 & 0.08 & 0.83 & 9.59 & 3 & 9.49 & 0.16 & 0.38 & 0.17 & 0.31\\
  134 & J233702.06+204756.0 & 0 & 0.0516 & 9.55 & 0.03 & -.-- & 0.70 & 6.93 & 2 & -.-- & -.-- & -.-- & -.-- & 0.00\\
  135 & J233703.24+205326.8 & 1 & 0.0576 & 9.04 & 0.06 & 0.39 & 1.00 & 1.73 & 2 & 9.43 & 0.84 & 0.43 & 0.21 & 0.26\\
  138 & J233703.72+205304.1 & 1 & 0.0580 & 9.48 & 0.04 & 0.01 & 0.54 & 3.44 & 1 & 9.74 & 0.46 & 0.34 & 0.17 & 0.47\\
  139 & J233705.27+210739.1 & 2 & 0.0638 & 9.38 & 0.05 & 0.05 & 0.36 & 1.30 & 2 & 9.27 & 1.27 & 0.32 & -.-- & 0.50\\
  142 & J233711.63+204243.3 & 2 & 0.0641 & 9.68 & 0.04 & -0.69 & 0.77 & 6.52 & 2 & 9.35 & 1.27 & 0.29 & -.-- & 0.72\\
  143 & J233712.13+204554.1 & 1 & 0.0581 & 9.15 & 0.11 & 0.02 & 0.88 & 2.86 & 2 & -.-- & -.-- & -.-- & -.-- & 0.34\\
  144 & J233712.73+210406.5 & 0 & 0.0584 & 9.35 & 0.05 & 0.31 & 0.48 & 1.90 & 1 & 9.68 & 0.18 & 0.84 & 0.32 & 0.54\\
  148 & J233720.02+204934.0 & 1 & 0.0583 & 9.17 & 0.07 & 0.12 & 0.88 & 4.31 & 2 & 9.45 & 0.22 & 0.86 & 0.32 & 0.70\\
  149 & J233720.15+213258.3 & 0 & 0.0577 & 9.31 & 0.08 & 0.28 & 0.68 & 3.57 & 2 & 9.67 & 1.27 & 0.25 & -.-- & 0.39\\
  150 & J233721.12+205717.6 & 0 & 0.0563 & 9.84 & 0.02 & 0.20 & 0.27 & 1.86 & 1 & 10.81 & 0.10 & 1.17 & 0.42 & 1.04\\
  153 & J233723.08+205719.7 & 0 & 0.0581 & 9.18 & 0.06 & 0.28 & 0.48 & 1.66 & 1 & -.-- & -.-- & -.-- & -.-- & 0.33\\
  157 & J233726.14+210016.1 & 0 & 0.0525 & 8.78 & 0.10 & -.-- & 0.98 & 2.70 & 1 & -.-- & -.-- & -.-- & -.-- & 0.42\\
  158 & J233731.24+205616.2 & 1 & 0.0575 & 9.26 & 0.09 & 0.00 & 0.29 & 1.61 & 2 & 8.63 & 1.28 & 0.31 & -.-- & 0.49\\
  159 & J233733.15+203213.4 & 1 & 0.0593 & 9.52 & 0.08 & -0.21 & 0.56 & 3.52 & 2 & 9.12 & 0.48 & 0.37 & 0.19 & 0.47\\
  160 & J233735.81+210101.8 & 0 & 0.0527 & 8.92 & 0.08 & 0.49 & 0.33 & 1.01 & 1 & 9.66 & 1.19 & 0.59 & 0.24 & 0.48\\
  161 & J233739.94+203118.8 & 1 & 0.0573 & 9.73 & 0.07 & 0.10 & 0.72 & 1.26 & 2 & 10.16 & 0.15 & 2.08 & 0.73 & 0.60\\
  162 & J233741.04+213051.6 & 0 & 0.0538 & 9.50 & 0.05 & 0.08 & 0.71 & 3.38 & 2 & 9.46 & 0.12 & 0.35 & 0.17 & 0.54\\
  166 & J233745.24+210742.1 & 2 & 0.0640 & 9.50 & 0.04 & 0.10 & 0.53 & 2.72 & 1 & 9.43 & 1.27 & 0.38 & -.-- & 0.39\\
  169 & J233746.51+212126.6 & 0 & 0.0533 & 9.07 & 0.06 & 0.33 & 0.62 & 3.37 & 2 & -.-- & -.-- & -.-- & -.-- & 0.38\\
  170 & J233748.75+204013.3 & 1 & 0.0561 & 9.64 & 0.04 & 0.15 & 0.38 & 2.82 & 1 & 10.12 & 0.12 & 2.27 & 0.79 & 0.53\\
  172 & J233750.60+212454.3 & 0 & 0.0553 & 9.32 & 0.06 & 0.09 & 0.35 & 0.53 & 2 & -.-- & -.-- & -.-- & -.-- & 0.29\\
  173 & J233751.32+211127.1 & 0 & 0.0553 & 9.58 & 0.04 & -0.04 & 0.36 & 1.49 & 2 & 9.11 & 1.27 & 0.24 & -.-- & 0.39\\
  176 & J233756.65+205436.8 & 1 & 0.0559 & 9.21 & 0.06 & 0.49 & 0.54 & 2.55 & 2 & 9.99 & 0.24 & 0.70 & 0.28 & 0.62\\
  177 & J233756.99+212234.8 & 0 & 0.0556 & 9.16 & 0.06 & -.-- & 0.97 & 6.03 & 2 & -.-- & -.-- & -.-- & -.-- & 0.26\\
  178 & J233758.90+204000.6 & 1 & 0.0551 & 9.60 & 0.05 & 0.13 & 0.72 & 4.04 & 2 & 9.60 & 0.14 & 0.4 & 0.20 & 0.60\\
  179 & J233800.62+205722.0 & 0 & 0.0559 & 9.52 & 0.05 & 0.55 & 0.58 & 2.00 & 2 & 10.72 & 0.10 & 5.07 & 1.77 & 0.79\\
  182 & J233808.40+205755.0 & 0 & 0.0551 & 9.39 & 0.06 & 0.19 & 1.00 & 12.62 & 3 & 9.98 & 0.32 & 0.38 & 0.19 & 0.60\\
  191 & J233825.56+204911.2 & 1 & 0.0564 & 9.86 & 0.03 & 0.05 & 0.56 & 3.09 & 1 & 9.81 & 0.16 & 0.54 & 0.21 & 0.52\\
  192 & J233826.00+203517.6 & 1 & 0.0577 & 9.83 & 0.06 & -0.10 & 0.39 & 1.42 & 2 & 9.71 & 1.27 & 0.51 & -.-- & 0.57\\
  194 & J233828.47+212848.0 & 0 & 0.0533 & 9.41 & 0.10 & 0.20 & 0.84 & 6.03 & 2 & 9.89 & 0.10 & 0.91 & 0.34 & 0.61\\
  196 & J233832.29+203644.9 & 1 & 0.0565 & 9.51 & 0.09 & 0.16 & 0.68 & 5.22 & 2 & 9.52 & 1.27 & 0.49 & -.-- & 0.34\\
  198 & J233841.11+212336.3 & 2 & 0.0633 & 9.57 & 0.07 & 0.16 & 1.00 & 17.94 & 3 & -.-- & -.-- & -.-- & -.-- & 0.86\\
  200 & J233846.64+204855.3 & 1 & 0.0555 & 9.31 & 0.07 & 0.09 & 0.67 & 6.26 & 2 & 8.96 & 1.27 & 0.23 & -.-- & 0.41\\
  201 & J233847.26+204720.0 & 1 & 0.0568 & 9.50 & 0.06 & 0.26 & 0.65 & 2.95 & 2 & 10.07 & 0.24 & 0.57 & 0.23 & 0.43\\
  204 & J233848.60+212311.0 & 2 & 0.0637 & 9.41 & 0.09 & -0.01 & 0.62 & 2.84 & 2 & 9.90 & 0.33 & 2.22 & 0.79 & 0.36\\
  205 & J233848.61+205550.2 & 1 & 0.0562 & 9.73 & 0.05 & -0.08 & 0.60 & 3.16 & 2 & 9.99 & 0.14 & 2.43 & 0.85 & 0.50\\
  210 & J233859.60+211723.6 & 1 & 0.0558 & 9.52 & 0.06 & 0.19 & 0.61 & 4.74 & 2 & 9.07 & 1.27 & 0.42 & -.-- & 0.49\\
  218 & J233948.55+205857.4 & 0 & 0.0565 & 9.88 & 0.09 & -0.10 & 2.00 & 1.42 & 1 & 9.90 & 0.16 & 1.26 & 0.45 & 0.44\\
\hline
\caption*{ Column (1): The assigned \hi identifier from  \citealt{HD2021}. Column (2): The SDSS identifier for the optical counterpart of the \hi detected galaxies, based on their Right Ascension and Declination (J2000.0). Column (3): Substructure identifier. SS=`0',`1': non-substructure and substructure galaxies in A2626, SS=`2': galaxies in the Swarm. Column (4): Redshift measured as the midpoint of the 20\% line width of the \hi global profile. Column (5,6): Log of \hi mass and uncertainty as mentioned in \citealt{HD2021}. Column (7): \hi deficiency using the scaling relation from \citet{denes2014}. Column (8): Calculated modified asymmetry using the definition of \citet{Lelli2014}. Column (9): Measured offset of the \hi centre from the optical centre (in kpc). Column (10): Visual classifications of galaxies. `1': settled sources, `2': disturbed sources, `3': unsettled sources. Column (11,12): Log of stellar mass and uncertainty derived from M/L from \citet{cluver2014}. Column (13,14): 12 $\mu$m star formation rates star formation rate and uncertainty from WISE observations (provided by T. Jarrett, private communication). Column (15): g-r magnitude from DECaLS \citep{dey2019} survey (corrected for extinction.)}

\label{tab:catalogue}
\end{longtable}

\end{appendix}
\end{document}